\documentstyle[11pt,seceq]{article}

 \newtheorem{Th}{Theorem}
 \newtheorem{Prop}{Proposition}
 \newcommand{\be}{\begin{equation}}
 \newcommand{\ee}{\end{equation}}
 \newcommand{\lb}{\label}
 
 \newcommand{\bh}{{\bf h}}
 \newcommand{\bm}{{\bf m}}
 \newcommand{\obm}{{\overline{{\bf m}}}}
 \newcommand{\bz}{{\bf 0}}
 \newcommand{\bD}{{\bf D}}
 \newcommand{\bF}{{\bf f}}
 
 \newcommand{\bZ}{{\bf z}}
 \newcommand{\BZ}{{\bf Z}}
 \newcommand{\hBZ}{\hat{{\bf Z}}}
 \newcommand{\HZ}{\hat{Z}}
 \newcommand{\obZ}{{\overline{{\bf z}}}}
 \newcommand{\oBZ}{{\overline{{\bf Z}}}}
 \newcommand{\bV}{{\bf V}}
 \newcommand{\obV}{{\overline{{\bf V}}}}
 \newcommand{\oV}{{\overline{V}}}
 \newcommand{\bK}{{\bf K}}
 \newcommand{\bA}{{\bf A}}
 
 \newcommand{\bB}{{\bf B}}
 \newcommand{\bL}{{\bf L}}
 
 \newcommand{\cH}{{\cal H}}
 \newcommand{\cP}{{\cal P}}
 \newcommand{\cD}{{\cal D}}
 \newcommand{\hL}{{\hat{L}}}
 \newcommand{\bx}{{\bf x}}
 \newcommand{\bX}{{\bf X}}
 \newcommand{\cZ}{{\cal Z}}
 \newcommand{\bcZ}{{\bf {\cal Z}}}
 \newcommand{\cN}{{\cal N}}
 \newcommand{\om}{{\overline{m}}}
 \newcommand{\oPsi}{{\overline{\Psi}}}
 \newcommand{\oalpha}{{\overline{\alpha}}}
 \newcommand{\balpha}{{\mbox{\boldmath $\alpha$}}}
 \newcommand{\obalpha}{\overline{\mbox{\boldmath $\alpha$}}}
 \newcommand{\bpi}{{\mbox{\boldmath $\pi$}}}
 \newcommand{\bpsi}{{\mbox{\boldmath $\psi$}}}
 \newcommand{\bdot}{{\mbox{\boldmath $\cdot$}}}

\begin{document}
 \title{Action Principle in Nonequilibrium Statistical Dynamics}
 \author{Gregory L. Eyink\\{\em Department of Mathematics, University of
Arizona}\\
	 {\em Building No. 89, Tucson, AZ, 85721}}
 \date{ }
 \maketitle
 \begin{abstract}
 We introduce a variational method for approximating distribution functions
 of dynamics with a ``Liouville operator'' $\hL,$ in terms of a {\em
nonequilibrium
 action functional} for two independent (left and right) trial states. The
 method is valid for deterministic or stochastic Markov dynamics, and for
stationary
 or time-dependent distributions. A practical Rayleigh-Ritz procedure is
advanced,
 whose inputs are finitely-parametrized ansatz for the trial states, leading to
a
 ``parametric action'' for their evolution. The Euler-Lagrange equations of the
 action principle are Hamiltonian in form (generally noncanonical). This
permits
 a simple identification of fixed points as critical points of the parametric
Hamiltonian.

 We also establish a variational principle for low-order statistics, such as
mean values
 and correlation functions, by means of {\em least effective action.} The
latter is a
 functional of the given variable, which is positive and convex as a
consequence of H\"{o}lder
 realizability inequalities. Its value measures the ``cost'' for a fluctuation
from the
 average to occur and in a weak-noise limit it reduces to the Onsager-Machlup
action. In
 general, the effective action is shown to arise from the nonequilibrium action
functional
 by a constrained variation. This result provides a Rayleigh-Ritz scheme for
calculating just
 the desired low-order statistics, with internal consistency checks less
demanding than
 for the full distribution.

 \end{abstract}

 \newpage

 \section{Introduction}

 The Rayleigh-Ritz variational method is a well-established technique in
quantum mechanics (e.g. see \cite{MF}).
 In this method one solves  approximately the stationary Schr\"{o}dinger's
equation by making a physically motivated
 trial {\em ansatz} for the ground-state wavefunction and then varying the
energy-expectation functional with respect
 to its parameters. A similar method is available for solving the
time-dependent Schr\"{o}dinger equation, based upon
 the Dirac-Frenkel dynamic variational principle \cite{LEM,KK,Dir,Fre}. These
methods are among the very few tools in
 the arsenal of theoretical physics able to assault systematically
strong-coupling problems of quantum dynamics. They
 are especially useful in quantum field theory and many-body theory, where
alternative numerical approaches are
 expensive or unfeasible. In some cases---such as the BCS theory of
superconductivity---the variational principle has
 been the stepping-stone to an exact solution of the problem.

 In our opinion, nonequilibrium statistical mechanics has been lacking a
variational principle of the
 same flexibility and scope as in quantum theory, capable of determining the
probability density function (PDF)
 for both the steady-state and also the time-dependent solution to the
initial-value problem. This is particularly
 true for problems such as high Reynolds number turbulence and large-scale
dynamics of multiphase fluids, where
 there is no small parameter in which to make a perturbation expansion or
asymptotic development and strong fluctuations
 dominate the phenomena on a wide range of length-scales. An obvious analogy
exists between Schr\"{o}dinger's equation
 for the wave-function and the {\em Liouville equation} for the PDF in the
nonequilibrium problems:
 \be \partial_t P=\hL P. \lb{LioEq} \ee
 This analogy has been used before to express classical statistical dynamics as
a formal quantum field theory
 in the work of Martin-Siggia-Rose (MSR) \cite{MSR}. It was noted in \cite{MSR}
that variational principles could be
 formulated, without any further details. However, a mathematical obstacle
exists to applying by analogy the quantum
 principles because the formal ``Hamiltonian'' $\hL$ is generally non-Hermitian
for the dissipative dynamical systems of
 interest. Variational methods of the standard form as in quantum mechanics
have been employed in special cases where
 $\hL$ can be transformed to Hermitian form \cite{Ris,MRV,BSS} or else based
upon the Hermitian squared operator
 $\hL^\dagger\hL$ \cite{Ris,Sey}. These methods seem to be either too
restrictive or too cumbersome to be as useful as
 the corresponding quantum principles. Recently, we have observed in the
turbulence context that a variational
 method may be developed for nonequilibrium dynamics which preserves the
principal advantages of the quantum method
 \cite{Eyi}. The key idea in the new formulation is to vary jointly over {\em
independent left and right trial states}.
 Although this Rayleigh-Ritz method seems to be most natural for a
non-Hermitian operator, it does not seem
 to have been previously used for nonequilibrium dynamics. It is our purpose
here to develop this method in a general
 context and in some formal detail.

 One advantage of the variational method in our formulation is that it yields,
by a procedure of {\em constrained
 variation}, a characterization of the {\em effective action} for any selected
statistic of interest, such as a mean-value
 or a two-point correlation. The effective action is a non-negative, convex
functional whose minimum is achieved
 by the true ensemble-average value. In quantum field theory the concept has it
roots in the early work of
 Heisenberg \& Euler \cite{HE} and Schwinger \cite{Sch} in QED.  In
nonequilibrium statistical mechanics, the
 first such action principle seems to have been Onsager's 1931 ``principle of
least dissipation'' \cite{O}, which applies
 to systems subject to thermal or molecular noise, governed by a
fluctuation-dissipation relation. A formulation of the
 least-dissipation principle by an action functional on histories was developed
in 1953 by Onsager and Machlup \cite{OM}.
 The effective action we consider coincides in a weak-noise limit with the
Onsager-Machlup action, as discussed some
 time ago by Graham \cite{Gra}. For vanishing noise, a path-integral formula
for the effective action can be evaluated
 by steepest descent, yielding the ``classical'' action of Onsager-Machlup.
However, in the strong-noise case,
 efficient calculational tools remain to be developed. We show here that the
Rayleigh-Ritz method provides one such
 computational scheme. The basis of this method is a generalization of
Symanzik's theorem in Euclidean field theory
 \cite{Sym} (see also \cite{CJT}), which characterizes the static effective
action, or, ``effective potential,'' by a
 constrained variation of the quantum energy-expectation functional. This
theorem has been extended by us to MSR field
 theory with non-Hermitian Hamiltonian operator \cite{Eyi}. Here we shall, for
completeness, briefly recapitulate that
 result and then expound in detail the corresponding Rayleigh-Ritz method. We
also establish a  Symanzik-type theorem
 for the time-dependent effective action, extending the earlier result of
Jackiw \& Kerman in quantum theory \cite{JK}
 to the initial-value problem in nonequilibrium statistical dynamics.

 The methods we develop here are quite general and apply, indeed, to the
solution of any large-scale
 stochastic system, not only those in nonequilibrium statistical physics, but
also to population dynamics in biology,
 to stochastic market models in mathematical finance, etc. The advantages of a
variational scheme are well-known.
 For example, we quote:
 \begin{quote}
 ``The great virtue of the variational treatment, `Ritz's method', is that it
permits efficient use in the process
 of calculation, of any experimental  or intuitive insight which one may
possess concerning the problem which is to be
 solved by calculation. It is important to realize that this is not possible,
or possible to a much smaller extent,
 if one performs the calculation by using the original form of the equations of
motion... Ritz's method, on the
 other hand, is definitely a method of successive approximations, and one which
converges better in the later
 stages of the approximation. Any information therefore which one may
possess---no matter whether it comes
 from experiments, from intuition, or from general experience obtained in
previous works on similar problems---
 can be made useful by using it in formulating the point of departure, the
`zeroeth approximation'.''
 (J. von Neumann, \cite{JvN}, p.357).
 \end{quote}
 The present paper elaborates the theoretical foundation of such a variational
scheme for stochastic dynamical
 systems. In future work we shall apply the method to various concrete systems
of practical interest. In particular,
 the paper \cite{AlEy1} demonstrates the feasibility of the Rayleigh-Ritz
method for numerical computation
 of the effective potential, and \cite{AlEy2} applies the action principle to
the problem of moment closures
 in turbulence modelling.

 \newpage

 \section{The Variational Method for Distributions}

 Our problem is to calculate the probability distribution functions (PDF's),
denoted by  $P,$
 for nonequilibrium Markov dynamics, governed by an equation of the form of
Eq.(\ref{LioEq}),
 where $\hL$ is the (forward) Markov generator. Concrete examples of practical
 interest are the nonequilibrium master equations \cite{vK}, and, as a
particular case,
 the Fokker-Planck equations \cite{Ris}, with
 \be \hL= -{{\partial}\over{\partial x_i}}\left(K_i(\bx)\cdot\right)
      +{{1}\over{2}}{{\partial^2}\over{\partial x_i\partial
x_j}}\left(D_{ij}(\bx)\cdot\right), \lb{FP} \ee
 in which $\bK$ is the drift vector and $\bD$ is the diffusion tensor. A
degenerate case
 of the latter of special interest occurs for zero noise ($\bD=\bz$), which is
 \be \hL= -{{\partial}\over{\partial x_i}}\left(K_i(\bx)\cdot\right), \lb{LO}
\ee
 the ``Liouville operator'' of the deterministic dynamical system
$\dot{\bx}=\bK(\bx).$

 We develop here a simple variational method to calculate approximately the
solutions
 of the Eq.(\ref{LioEq}) for $P$, both for the stationary PDF, $P_s,$ and for
time-dependent solutions $P_t$
 with prescribed initial data $P_0.$ Our methods are analogous to Rayleigh-Ritz
procedures
 traditional in quantum mechanics, but with a modification due to the fact that
the operator
 $\hL$ is non-self-adjoint:
 \be \hL^\dagger\neq \hL. \lb{notSA} \ee
 Although the spectra of $\hL$ and $\hL^\dagger$ are the same (because $\hL$ is
a real operator: $\hL^*=\hL$),
 their eigenstates are distinct. Equivalently, the left and right eigenstates
of $\hL$ are distinct \cite{MF,Wk}.
 This is particularly true for the ``ground states''
 \be \hL|\Omega^R\rangle=0 \,\,\,\,\,\,\,\,\hL^\dagger|\Omega^L\rangle=0.
\lb{GS} \ee
 Because of the fundamental asymmetry of the problem, Hilbert space or $L^2$
methods are not so useful
 as in quantum theory. Instead, the standard mathematical formulation (see
\cite{GS}) is to take
 $\hL$ as an operator on $L^1,$ considered as a space of ``normalizable
states,'' and $\hL^\dagger$
 as an operator on $L^\infty,$ considered as a space of ``bounded
observables.''
 \footnote{The mathematical notation is, unfortunately, the opposite to that
generally adopted in
 the physics literature: what we have called $\hL,\hL^\dagger$ are in
mathematics usually denoted as $L^*, L,$ (forward
 and backward Markov operators, respectively).} Although the inequality of the
two ground states is a complication,
 there are special features that largely compensate for this. The ``right
ground state'' $\Omega^R$
 is the main unknown of the problem, the stationary PDF, $P_s,$ and it can
always be taken to be non-negative
 \be \Omega^R(\bx)\geq 0. \lb{pos} \ee
 This is part of the statement of the Perron-Frobenius theorem, since $e^{tL}$
is an operator with strictly positive
 kernel: see \cite{Kras}, or Theorem 3.3.2 of \cite{GJ}. On the other hand, the
``left ground state'' is known
 {\em exactly} a priori:
 \be \Omega^L(\bx)\equiv 1. \lb{LGS} \ee
 This latter fact turns out to be of great utility in our method. We discuss
first the stationary problem
 and thereafter consider the time-dependent case.

 \noindent {\em (i) Stationary Distributions}

 Define a functional ${\cal H}$ of left and right state vectors, as
 \be {\cal H}[\Psi^R,\Psi^L]\equiv \langle\Psi^L,\hL\Psi^R\rangle. \lb{QH} \ee
 Then it is easy to see that $\Omega^R,\Omega^L$ are uniquely characterized as
the
 joint extremal point of the functional $\cH$:
 \be \delta{\cal H}[\Psi^R,\Psi^L]=0\,\,\,\leftrightarrow\,\,\,(\Psi^R,\Psi^L)
=(\Omega^R,\Omega^L). \lb{Vc} \ee
 In fact,
 \be \delta{\cal H}[\Psi^R,\Psi^L]=\langle\delta\Psi^L,\hL\Psi^R\rangle
+\langle\Psi^L,\hL\cdot\delta\Psi^R\rangle=0,  \lb{vQH} \ee
 if and only if
 \be \hL|\Psi^R\rangle=0\,\,\,\,\&\,\,\,\,\hL^\dagger|\Psi^L\rangle=0. \lb{gsc}
\ee
 As stated above, we take $\Psi^R\in L^1$ (``states'') and $\Psi^L\in L^\infty$
(``observables''), with
 \be \langle\Psi^L,\Psi^R\rangle\equiv\int d\bx\,\,\Psi^L(\bx)^*\Psi^R(\bx).
\lb{braket} \ee
 The ``inner product'' notation is always used in this paper as the canonical
sesquilinear association of $\Psi^R\in L^1$
 and $\Psi^L\in L^\infty$ with the complex number $\langle\Psi^L,\Psi^R\rangle$
defined in Eq.(\ref{braket}).

 This simple variational characterization of the ground states can be made the
basis of a
 Rayleigh-Ritz method of approximation. To initiate this method, one must make
{\em trial ansatz}
 \be \Psi^R=\Psi^R(\balpha)\,\,\,\,\&\,\,\,\,\Psi^L=\Psi^L(\balpha), \lb{ans}
\ee
 for the ground states. \footnote{Since we know $\Omega^L$ to  be exactly equal
to one, it may seem unnecessary to make
 an ansatz for it at all. However, variation over the  ``observables'' is
required to characterize the ``state,'' or
 right ground-state $\Omega^R.$} The vector $\balpha=(\alpha_1,...,\alpha_N)$
denotes a set of $N$ real parameters
 (where possibly $N=\infty$). In certain cases, we shall wish to have
dependence of some parameters only in one
 of the vectors $\Psi^H,\,H=L,R$ and we denote the corresponding parameters as
$\balpha^H=
 (\alpha_1^H,...,\alpha^H_{N^H}),$ for $H=L,R$ respectively. We then use
$\balpha$ to denote only the
 common parameters in both trial vectors. An interesting special case is when
there are no such
 common parameters, i.e.
 \be \Psi^R=\Psi^R(\balpha^R)\,\,\,\,\&\,\,\,\,\Psi^L=\Psi^L(\balpha^L),
\lb{ansII} \ee
 and $N^L=N^R,$ i.e. with equal numbers of the left- and right-parameters. The
ansatz
 provide an explicit, but arbitrary, reduction of the original variational
problem in an infinite-dimensional
 function space to an analogous problem in $N$-dimensional Euclidean space. A
given assumed form of the trial ansatz
 provides, in essence, a ``nonlinear projection'' of the original
time-independent stationarity equations. This is
 the same general strategy proposed explicitly by Bayly under the term
``parametric PDF closures'' \cite{Bay}
 (and used implicitly by others before). Here we simply explain how this
strategy may be implemented variationally.

 For any particular ansatz, we denote
 \be \cH(\balpha)\equiv \langle\Psi^L(\balpha),\hL\Psi^R(\balpha)\rangle,
\lb{qH} \ee
 which we call the {\em (parametric) Hamiltonian}. We may now seek for the
extremal, or critical, points of $\cH$:
 \be {{\partial\cH}\over{\partial\alpha_i}}(\balpha_*)=0. \lb{stveq} \ee
 This condition may be written more explicitly as
 \be \langle\psi^L_i(\balpha_*),\hL\Psi^R(\balpha_*)\rangle
 +\langle\Psi^L(\balpha_*),\hL\psi^R_i(\balpha_*)\rangle=0  \lb{stveq2}
 \ee
 for each $i=1,...,N,$ where, in general, for $H=L,R$
 \be \psi_i^H(\balpha)={{\partial\Psi^H}\over{\partial\alpha_i}}(\balpha).
\lb{MF} \ee
 One may take the corresponding state vectors as the approximations to the
ground states:
 \be \Omega^R_*(\bx)=\Psi^R(\bx;\balpha_*)\,\,\,\,\&\,\,\,\,
              \Omega^L_*(\bx)=\Psi^L(\bx;\balpha_*). \lb{apgs} \ee
 In the special case Eq.(\ref{ansII}) with no common parameters, the
variational equations become simply
 \be  0={{\partial\cH}\over{\partial\alpha_i^R}}=
		 \langle\Psi^L(\balpha^L_*),\hL\psi^R_i(\balpha^R_*)\rangle \lb{stveqIIR} \ee
 and
 \be  0={{\partial\cH}\over{\partial\alpha_i^L}}=
		 \langle\psi^L_i(\balpha^L_*),\hL\Psi^R(\balpha^R_*)\rangle, \lb{stveqIIL}
\ee
 with $i=1,...,N(=N^R=N^L).$ We may also write out the general
Eq.(\ref{stveq2}) more explicitly as separate equations
 for the variations under each of $\balpha^R,\balpha^L,$ and $\balpha.$
However, we have not found this
 version of the equations to be as useful, so that we relegate it to an
Appendix.

 In general, the function $\cH(\balpha)$ may have more than one critical point.
 Some {\em a priori} criteria for selection of the critical point(s) of
interest arise from the
 {\em exact} information for the problem that $\cH[\Omega^R,\Omega^L]=0$ and
that $\Omega^L\equiv 1.$
 Hence, among the possible critical points, we should only accept those for
which
 \be \cH(\balpha_*)\approx 0, \lb{zC} \ee
 and
 \be \Psi^L(\bx;\balpha_*)\approx 1. \lb{lgsc} \ee
 The second condition generally implies the first.  Hence, we should only
accept those critical
 points for which $\Omega^L_*$ is close to the  constant 1. We refer to such
critical points as ``acceptable.''
 Because of the acceptability condition, we see that the ansatz need really
only explore the region near
 $\Psi^L\approx 1.$  Thus, we may without loss of generality assume that
$\alpha_i^L\ll 1$ and expand to linear order:
 \be  \Psi^L(\balpha,\balpha^L)= 1+\sum_{i=1}^{N^L}\alpha_i^L\psi_i^L(\balpha),
\lb{ansl} \ee
 where, now, for
$H=L,R,\,\,\psi_i^H(\balpha,\balpha^H)
={{\partial\Psi^H}\over{\partial\alpha_i^H}}(\balpha,\balpha^H),$
 rather than Eq.(\ref{MF}).  Correspondingly,
 \be \cH(\balpha^R,\balpha^L,\balpha)=\alpha^L_i\langle\psi_i(\balpha),
              \hL\Psi^R(\balpha,\balpha^R)\rangle \lb{qHlin} \ee
 (summation convention implied). \footnote{To guarantee $\Psi^L\in L^\infty,$
we should really take
 \be  \Psi^L(\alpha,\alpha^L)= \exp\left[
i\sum_{i=1}^{N^L}\alpha_i^L\psi_i^L(\alpha)\right]. \lb{ansexp} \ee
 However, this leads to equivalent results as Eq.(\ref{ansl}).}

 It is useful to consider the special case Eq.(\ref{ansII}) with no common
parameters,
 for which the variational Eqs. (\ref{stveqIIR}), (\ref{stveqIIL}) become
simply
 \be \alpha_i^L\langle\psi_i^L,\hL\psi_j^R(\balpha^R)\rangle=0, \lb{stveqIIR'}
\ee
 and
 \be \langle\psi_i^L,\hL\Psi^R(\balpha^R)\rangle=0. \lb{stveqIIL'} \ee
 for $i,j=1,...,N.$ If the matrix in Eq.(\ref{stveqIIR'}) is nonsingular,
 \be  {\rm det}\left[\langle\psi_i^L,\hL\psi_j^R(\balpha^R)\rangle\right]\neq
0, \lb{nonsing} \ee
 then the first of the variational equations has as its {\em unique} solution
 \be \balpha^L_*\equiv \bz. \lb{zerosln} \ee
 In that case, Eq.(\ref{stveqIIL'}) is the only remaining equation and it
determines the critical value $\balpha^R_*.$
 Thus, the condition determining $P_s=\Omega^R$ in this approximation is the
stationarity condition
 \be \langle \hL^\dagger\psi^L_i\rangle_{\balpha^R_*}=0, \lb{stmc} \ee
 for the finite set of moment-functions $\psi_i^L,i=1,...,N^L.$ In that case,
the variational method does not
 differ from the projection of the dynamics onto a finite set of moments. If
one permits a more general dependence
 of $\Psi^L$ on the parameters $\balpha^L$ than the linear ansatz
Eq.(\ref{ansl}), then the variational
 method does not generally coincide with moment projection. However, we see no
advantage at this point
 to allowing a nonlinear dependence on $\balpha^L.$

 It is possible to obtain the moment projection condition in a slightly more
general form, i.e. so that the moments
 $\psi_i^L$ depend upon the same set of parameters $\balpha$ as the trial state
$\Psi^R=\Psi^R(\balpha).$
 Formally, we take $N^R=0, N=N^L.$ We may obtain for the $N$ parameters
$\balpha$ determining equations of the form
 \be \langle\psi_i^L(\balpha),\hL\Psi_j^R(\balpha)\rangle=0, \lb{stveqIIL*} \ee
 or, equivalently,
 \be \langle \hL^\dagger\psi^L_i(\balpha)\rangle_{\balpha}=0. \lb{stmc*} \ee
 This is accomplished by making the variational ansatz
 \be \Psi^L= 1+\sum_{i=1}^N\alpha_i^L\psi_i^L(\balpha)
 \,\,\,\,\&\,\,\,\,\Psi^R=\Psi^R(\balpha). \lb{ansl*} \ee
 There may be some advantage in permitting the moments to vary along with the
trial state. Hence, this
 more general version is worked out in the Appendix.

A simple example of such ansatz as discussed above may be devised based upon a
{\em trial weight} $w=w(\bx),$
which is a normalized probability density, and an adapted set of {\em
orthogonal polynomials} $p_n(\bx)$:
\be \int d\bx\,w(\bx)p_n(\bx)p_{n'}(\bx)=\delta_{nn'}. \lb{orplc} \ee
See \cite{Kr70,Sz}. A natural form of the trial ansatz then takes
$N^R=N^L(\equiv N)$ and
\be \Psi^R(\bx;\balpha^R)= w(\bx)\cdot\sum_{n=0}^{N-1}\alpha^R_n p_n(\bx),
\lb{oprans} \ee
and
\be \Psi^L(\bx;\balpha^L)= \sum_{n=0}^{N-1}\alpha^L_n p_n(\bx). \lb{oplans} \ee
This ansatz is a simple case of the type of Eq.(\ref{ansII}), with no common
parameters. Here the stationarity condition
becomes simply
\be \bL_N\balpha^R_*=0\,\,\,\,\&\,\,\,\,\balpha^L_*\bL_N=0, \lb{opstc} \ee
with
\be \left(\bL_N\right)_{nn'}\equiv \langle p_n,\hL(w\cdot p_{n'})\rangle.
\lb{defL} \ee
for $0\leq n,n'<N.$ In other words, the $\balpha^R_*$ and $\balpha^L_*$ should
be, respectively, right and left
eigenvectors of the matrix $\bL_N$ with eigenvalue zero. It is easy to check
that a left eigenvector of $\bL_N$
for the eigenvalue zero always exists and is given simply by
\be \alpha^L_{*n}=\delta_{n,0}. \lb{lefteigvc} \ee
It is possible to generalize the orthogonal polynomial ansatz by choosing the
trial weight $w(\balpha),$
depending upon some additional $M$ parameters $\alpha_i,\,\,i=1,...,M.$ In that
case, the adapted orthogonal
polynomials will depend also upon $\balpha.$ After initial variation over
$\balpha^R,\balpha^L,$ a second
variation may be made to optimize the choice of $\balpha.$

An advantage of the orthogonal polynomial scheme is that it may converge in the
limit $N\rightarrow \infty$:
for an example, see \cite{AlEy1}. Some sufficient conditions for convergence
are discussed in \cite{Eyi}. It is
necessary for convergence that
$\int d\bx\,{{P_s^2(\bx)}\over{w(\bx)}}<\infty$ \cite{Kr70}. Unfortunately, the
expansion ansatz Eq.(\ref{oprans})
for the state need not be positive at all values of $\bf x$. Instead, {\em
realizability} can be guaranteed
by making an ansatz
\be \Psi^R=w(\balpha,\balpha^R), \lb{wans} \ee
in which
\be w(\bx;\balpha,\balpha^R)\geq 0,\,\,\,\,\,\,\,\int
d\bx\,w(\bx;\balpha,\balpha^R)=1. \lb{wansc} \ee
This assures realizability whenever such an ansatz, along with Eq.(\ref{ansl}),
yields an ``acceptable''
critical point. The criterion of realizability is especially important for a
few parameter ansatz,
incorporating certain physical insights and ideas, as a test of those beliefs.
On the other hand, for the case where
$N\rightarrow\infty,$ it may be preferable to impose the criterion of {\em
convergence}. This might be
done even at the price of loss of realizability, if convergence for a statistic
of particular interest
is rapid enough. The dual criteria of realizability and convergence ought to be
regarded as complementary in their
applicability.

\noindent{\em (ii) Time-Dependent Distributions}

We first observe how the evolution equation Eq.(\ref{LioEq}) may be formulated
variationally. Let us define
\be \Gamma[\Psi^R,\Psi^L]\equiv \int_0^\infty
dt\,\,\langle\Psi^L(t),(\partial_t-\hL)\Psi^R(t)\rangle, \lb{act} \ee
as a functional of ``trajectories'' $\Psi^H(t),\,H=L,R.$ We refer to this
functional as the {\em nonequilibrium action}.
It is easy to see formally that the stationarity condition
\be \delta\Gamma[\Psi^R,\Psi^L]=0, \lb{stact} \ee
is equivalent to
\be (\partial_t-\hL)|\Psi^R(t)\rangle=0\,\,\,\,\&\,\,\,\,
      (\partial_t+\hL^\dagger)|\Psi^L(t)\rangle=0, \lb{dvareq} \ee
the variation being performed with the constraint
\be
\langle\Psi^L(\infty),\Psi^R(\infty)\rangle=\langle\Psi^L(0),\Psi^R(0)\rangle.
\lb{endpt} \ee
In other words, a pair of trajectories is an extremal point of the action if
and only if the ``right trajectory''
is a solution of the evolution equation Eq.(\ref{LioEq}) and the ``left
trajectory'' is a solution of the adjoint
equation, subject to the ``endpoint constraint'' Eq.(\ref{endpt}). It is
important to note a particular exact
solution of the adjoint equation
\be \Psi^L(\bx,t)\equiv 1. \lb{unit} \ee
In that case, the endpoint constraint becomes
\be \int d\bx\,\,\Psi^R(\bx,\infty)=\int d\bx\,\,\Psi^R(\bx,0), \lb{1endpt} \ee
which is automatically satisfied by any solution of the evolution equation. In
other words, $\Psi^L(t)\equiv 1$
together with any solution $\Psi^R(t)$ of the evolution equation provides an
extremal point of the action
$\Gamma[\Psi^R,\Psi^L].$ In this important special case
$\Gamma[\Psi^R,\Psi^L]=0.$ We may note the equivalent form
of the nonequilibrium action
\be \Gamma[\Psi^R,\Psi^L]\equiv \int_0^\infty
dt\,\,\left(\langle\Psi^L(t),\dot{\Psi}^R(t)\rangle
-\cH[\Psi^R(t),\Psi^L(t)]\right), \lb{Hact} \ee
which shows that $\Psi^L$ is formally a momentum $\Pi^R$ canonically conjugate
to $\Psi^R.$ In that case,
the evolution equation and its adjoint are formally restated as ``Hamilton's
equations''
\be \dot{\Psi}^R(\bx)={{\delta}\over{\delta\Psi^L(\bx)}}
\cH[\Psi^R,\Psi^L]\,\,\,\,\&\,\,\,\,
    \dot{\Psi}^L(\bx)= -{{\delta}\over{\delta\Psi^R(\bx)}}\cH[\Psi^R,\Psi^L].
\lb{Hameq} \ee
This makes it obvious that the Hamiltonian is invariant along an extremal set
of trajectories of the action
Eq.(\ref{Hact}).

In the same manner as for the stationary case, we may use the previous
variational principle as the basis
of an approximation method for the time-dependent PDF. The basic idea is
similar to time-dependent variational
principles of standard use in quantum mechanics \cite{LEM,KK}, going back to
the early work of Dirac \cite{Dir}
and Frenkel \cite{Fre}. The procedure is initiated by making trial ansatz for
the trajectories, in the form
\be \Psi^H(t)=\Psi^H(\balpha(t)) \lb{tdans} \ee
with $H=L,R.$ In other words, the reduction to finite number of degrees of
freedom is made with the same functional form
as for the stationary case and all of the time dependence is contained in the
parameters $\balpha(t).$
This is the same idea as in the general method of parametric PDF closure,
except that we here
derive equations for the closure parameters variationally. Indeed, we may
substitute the trial trajectories
into the action to obtain a reduced or {\em parametric action}
\be \Gamma[\balpha]\equiv \int_0^\infty
dt\,[\pi_i(\balpha(t))\dot{\alpha}_i(t)-\cH(\balpha(t))], \lb{paract} \ee
with
\be \pi_i(\balpha)\equiv \langle\Psi^L(\balpha),{{\partial}\over{\partial
\alpha_i}}\Psi^R(\balpha)\rangle. \lb{parmom} \ee
The Euler-Lagrange equations of the variational principle have the special
form:
\be \{\alpha_i,\alpha_j\}\dot{\alpha}_j=
{{\partial\cH}\over{\partial\alpha_i}}, \lb{ELeq} \ee
in which
\be \{\alpha_i,\alpha_j\}\equiv \langle{{\partial\Psi^L}\over{\partial
\alpha_i}}(\balpha),
                                       {{\partial\Psi^R}\over{\partial
\alpha_j}}(\balpha)\rangle
                               -\langle{{\partial\Psi^L}\over{\partial
\alpha_j}}(\balpha),
                                       {{\partial\Psi^R}\over{\partial
\alpha_i}}(\balpha)\rangle. \lb{Lbra} \ee
This is an infinite-dimensional generalization of the {\em Lagrange bracket} of
classical mechanics; see \cite{KK}
and \cite{Gold}, p.250. It is easily checked to have the properties
\be \{\alpha_j,\alpha_i\} = -\{\alpha_i,\alpha_j\}, \lb{anti} \ee
and
\be {{\partial}\over{\partial\alpha_i}}\{\alpha_j,\alpha_k\}
              +{{\partial}\over{\partial\alpha_j}}\{\alpha_k,\alpha_i\}
+{{\partial}\over{\partial\alpha_k}}\{\alpha_i,\alpha_j\}=0. \lb{Ljacobi} \ee
Let us first verify the stated form of the Euler-Lagrange equations
Eq.(\ref{ELeq}). The verification follows from
the result that
\be {{\delta}\over{\delta\alpha_i}}\int
dt\,\pi_i(\balpha)\dot{\alpha}_i=\{\alpha_i,\alpha_j\}\dot{\alpha}_j. \lb{prop}
\ee
By a simple calculation
\be {{\delta}\over{\delta\alpha_i}}\int dt\,\pi_i(\balpha)\dot{\alpha}_i
         =\langle{{\partial\Psi^L}\over{\partial
\alpha_i}},{{\partial\Psi^R}\over{\partial \alpha_j}}\rangle\dot{\alpha}_j
         +\langle\Psi^L,{{\partial^2\Psi^R}\over{\partial \alpha_i\partial
\alpha_j}}\rangle\dot{\alpha}_j
         -{{d}\over{dt}}\pi_i(\balpha). \lb{simp} \ee
However,
\be {{d}\over{dt}}\pi_i(\balpha)=
        \langle{{\partial\Psi^L}\over{\partial
\alpha_j}},{{\partial\Psi^R}\over{\partial \alpha_i}}\rangle\dot{\alpha}_j
         +\langle\Psi^L,{{\partial^2\Psi^R}\over{\partial \alpha_i\partial
\alpha_j}}\rangle\dot{\alpha}_j. \lb{dotpi} \ee
This yields Eq.(\ref{prop}). The property Eq.(\ref{anti}) of Lagrange
brackets is obvious. Eq.(\ref{Ljacobi})
follows from the expression Eq.(\ref{Lbra}) by a simple calculation.

If the matrix of Lagrange brackets $(\{\alpha_i,\alpha_j\})$ is non-degenerate,
that is, ${\rm det}
\left(\{\alpha_i,\alpha_j\}\right)\neq 0,$ then we may introduce a
corresponding {\em Poisson bracket}
$[\alpha_i,\alpha_j]$ as the elements
of the inverse matrix:
\be \left([\alpha_i,\alpha_j]\right)=\left(\{\alpha_i,\alpha_j\}\right)^{-1}.
\lb{defPbra} \ee
It is straightforward to show that the Poisson bracket has properties implied
by those of the Lagrange
bracket, Eqs.(\ref{anti}),(\ref{Ljacobi}), namely:
\be [\alpha_j,\alpha_i] = -[\alpha_i,\alpha_j], \lb{Panti} \ee
and
\be [\alpha_i,[\alpha_j,\alpha_k]]+
[\alpha_j,[\alpha_k,\alpha_i]]+[\alpha_k,[\alpha_i,\alpha_j]]=0. \lb{Pjacobi}
\ee
The latter is the well-known Jacobi identity. The bracket may be extended to
arbitrary functions $f$ and $g$
of coordinates $\balpha$ via the definition
\be [f,g]\equiv \sum_{p,q}\,\,{{\partial f}\over{\partial\alpha_p}}
                                  {{\partial
g}\over{\partial\alpha_q}}[\alpha_p,\alpha_q]. \lb{genPbra} \ee
With this definition, the Poisson bracket satisfies Eqs.(\ref{Panti}) and
(\ref{Pjacobi}) for all functions.
Note that the Jacobi identity for general functions follows by the argument of
\cite{Gold}, p.257.
The parametric equations may then be written as
\be \dot{\alpha}_i= [\alpha_i,\cH], \lb{parHameq} \ee
which are in Hamiltonian form. In general canonically conjugate variables do
not exist for this Hamiltonian
(i.e. the system is noncanonical Hamiltonian).  Notice that the Poisson
brackets $[\alpha_i,\alpha_j]$ of the
system depend only upon the parametrization (i.e. the trial ansatz) and that
the dynamics enters solely through
the Hamiltonian $\cH(\balpha).$  We now see very simply that the fixed points
of the parametric evolution equations
coincide with the critical points of the corresponding Hamiltonian.
\footnote{Even without the non-degeneracy condition
the fixed points would include all of the critical points of $\cH,$ although
there might be additional
fixed points.} Furthermore, the parametric Hamiltonian is an integral of motion
for the evolution equations.
Notice that, if the non-degeneracy condition failed at finite time, then the
solutions themselves to the parametric
equations might become ill-defined.

A case of special interest is that in which $\Psi^H=\Psi^H(\balpha^H),\,H=L,R,$
with an equal number of $\balpha^R$ and
$\balpha^L$ parameters. Observe that the Lagrange brackets are now given simply
as
\be \{\alpha_i^L,\alpha_j^R\}=
\langle\psi_i^L(\balpha^L),\psi_j^R(\balpha^R)\rangle \lb{lagbraLR} \ee
and
\be \{\alpha_i^R,\alpha_j^L\}=
-\langle\psi_j^L(\balpha^L),\psi_i^R(\balpha^R)\rangle \lb{lagbraRL}. \ee
with all other brackets vanishing. It is easy to check that the variables
$\bpi^R$ introduced as
\be \pi_i^R(\balpha_R,\balpha_L)\equiv
                \langle\Psi^L(\balpha^L),{{\partial}\over{\partial
\alpha_i^R}}\Psi^R(\balpha^R)\rangle, \lb{parmomR} \ee
satisfy
\be [\alpha_i^R,\pi_j^R]=\delta_{ij}, \lb{canconj} \ee
that is, $\bpi^R$ is the momentum canonically conjugate to $\balpha^R.$ If
$\pi_i^R(\balpha^R,\balpha^L)=\pi^R_i$
is invertible at each fixed $\balpha^R$ for $\balpha^L$ in terms of $\bpi^R$
and $\balpha^R,$ then by a change of
variables the system has canonical Hamiltonian form.

As in the static case, there is a criterion of ``acceptability''of solutions,
which requires that
$\Psi^L(t)\approx 1$ for all time $t.$ Let us consider first for simplicity the
previous special case with
$\Psi^H=\Psi^H(\balpha^H),\,H=L,R$. Just as for the statics, we are motivated
to adopt the linear ansatz
\be  \Psi^L(\bx;\balpha^L)= 1+\sum_{i=1}^{N}\alpha_i^L\psi_i^L(\bx).
\lb{anslII} \ee
In this case, the equations for $\balpha^L(t)$ become:
\be -\langle\psi_j^L,\psi_i^R(\balpha^R)\rangle\dot{\alpha}_j^L
         =\alpha_j^L\langle\psi_j^L,\hL\psi_i^R(\balpha^R)\rangle,
\lb{sparmeqR'} \ee
$i=1,...,N,$ which have as an {\em exact solution}
\be \balpha^L(t)\equiv \bz. \lb{exsoln} \ee
Within this same ansatz the equation remaining to be solved for $\balpha^R(t)$
reduces to:
\be \langle\psi_i^L,\psi_j^R(\balpha^R)\rangle\dot{\alpha}_j^R
                =\langle\psi_i^L,\hL\Psi^R(\balpha^R)\rangle.
\lb{sparmeqL'} \ee
For this case, a further simplification is possible by introducing {\em
moment-averages}
\be m_i(\balpha^R)\equiv \langle\psi_i^L\rangle_{\balpha^R}, \lb{moments} \ee
and the {\em dynamical vector}
\be V_i(\balpha^R)\equiv \langle\hL^\dagger\psi_i^L\rangle_{\balpha^R}.
\lb{dynvec} \ee
Because $\{\alpha^L,\alpha^R\}={{\partial m_i}\over{\partial
\alpha_j^R}}(\balpha^R)$ for the ansatz Eq.(\ref{anslII}),
it follows that
\be \{\alpha_i^L,\alpha^R_j\}\dot{\alpha}^R_j={{\partial m_i}\over{\partial
\alpha_j^R}}\dot{\alpha}_j^R=
                \dot{m}_i. \lb{mdot} \ee
Therefore, the equation of motion Eq.(\ref{sparmeqL'}) expressed in terms of
the moments $\bm$ becomes simply
\be \dot{m}_i= V_i(\bm), \lb{momeq} \ee
where $\bV(\bm)\equiv \bV(\balpha(\bm)).$ In this way we see how
``moment-closures'' as they have been
traditionally employed in nonequilibrium dynamics are obtained in our scheme.
Closure is achieved by
calculating all averages with respect to the PDF ansatz
$P(\bx,t)=\Psi^R(\bx;\balpha^R(t))$ and then
eliminating the parameters $\balpha^R(t)$ in terms of the (equal number of)
moments $\bm(t).$ As we
shall discuss in the next section, this variational method of moment-closure
has definite theoretical
advantages.

More generally, we may employ the ansatz Eq.(\ref{ansl*}),
$\Psi^L=1+\sum_{i=1}^N \alpha^L_i\psi_i^L(\balpha)
\,\,\,\,\&\,\,\,\,\Psi^R=\Psi^R(\balpha),$ allowing for some parameter
dependence of the moment-functions
$\psi_i^L(\balpha).$ This choice is considered in the Appendix, so we here just
report the results.
As with the case previously considered, it is not hard to check that
$\balpha^L(t)\equiv \bz$ is an
{\em exact solution} of its equation. The remaining equation for $\balpha$
takes the form
\be \{\alpha_i^L,\alpha_j\}\dot{\alpha}_j=V_i(\balpha) \lb{sparmeqL*} \ee
with
\be V_i(\balpha)\equiv \langle\hL^\dagger\psi_i^L(\balpha)\rangle_{\balpha},
\lb{gendynvec} \ee
generalizing Eq.(\ref{dynvec}), and
\be \{\alpha_i^L,\alpha_j\}=\langle\psi_i^L(\balpha),
             {{\partial\Psi^R}\over{\partial\alpha_j}}(\balpha)\rangle.
\lb{lagbraLG} \ee
By an easy calculation one can see also that
\be \{\alpha_i^L,\alpha_j\}={{\partial}\over{\partial\alpha_j}}
                           \langle\psi_i^L(\balpha)\rangle_\balpha
-\langle{{\partial\psi_i^L}\over{\partial\alpha_j}}(\balpha)\rangle_\balpha.
\lb{lagbraBB} \ee
Comparison with Eq.(2.11) in the work of Bayly \cite{Bay} reveals that the
Eq.(\ref{sparmeqL*})
obtained via the ansatz Eq.(\ref{ansl*}) is equivalent to the dynamical
equations obtained by ``moment-projection''
in the parametric PDF closure scheme. Here these equations are simply shown to
have a variational formulation.

As in the static case, a useful ansatz is provided by a fixed trial weight
$w(\bx)$ and orthogonal expansions
\be \Psi^R(\balpha^R)= w\cdot\sum_{n=0}^{N-1}\alpha^R_n p_n, \lb{fwopansR} \ee
and
\be \Psi^L(\balpha^L)= \sum_{n=0}^{N-1}\alpha^L_n p_n. \lb{fwopansL} \ee
In that case it is easy to calculate that
\be \{\alpha_n^L,\alpha^R_m\}= -\{\alpha_m^R,\alpha^L_n\}=\delta_{nm},
\lb{delbra} \ee
and that
\be \cH(\balpha^R,\balpha^L)=\sum_{n,m}\alpha^L_n (\bL_N)_{nm}\alpha^R_m.
\lb{quadH} \ee
Therefore we see that $\balpha^R$ and $\bpi^R=\balpha^L$ are canonically
conjugate, and the
parametric action is a quadratic form
\be \Gamma[\balpha^R,\balpha^L]=\int dt\,[\balpha^L\bdot\dot{\balpha}^R
-\balpha^L\bdot\bL_N{\balpha}^R]. \lb{quadact} \ee
In consequence, the evolution equations are linear
\be \dot{\balpha}^R=\bL_N\balpha^R \lb{lineqR}
\,\,\,\,\,\&\,\,\,\,\,\dot{\balpha}^L=-\balpha^L\bL_N, \lb{lineq} \ee
for this particular ansatz. The second equation has exact solution
$\alpha^L_n(t)\equiv \delta_{n0}.$ The first
equation is a standard Galerkin truncation of the linear Liouville dynamics,
Eq.(\ref{LioEq}).

\newpage

\section{Constrained Variation and Effective Action}

\noindent {\em (i) The Principle of Least Effective Action}

For spatially extended systems, or for any system with large numbers of degrees
of freedom, it is certainly
too ambitious to try to calculate the full PDF. Such a calculation would put
any trial ansatz
to an extremely severe test and could hardly be expected to succeed, in
general, with a few number of
parameters. In any case, the physical interest is usually in some special
low-order statistic,
such as a mean field or a correlation function. Such quantities are represented
by random variables
$\bZ$ on the microscopic phase space, that is, by functions $\bZ=\bZ(\bx)$ of
the dynamical
variables $\bx$. In practice one will be mostly interested in some simple
low-order moments of
the dynamical variables $\bx$ themselves, e.g. $\bZ=\bx, \bx\otimes\bx$, etc.
It should be possible to successfully
calculate a statistic of this type with a simpler ansatz with just a few
parameters, if those are insightfully
chosen. However, the variational method, as we have described it so far, allows
one to calculate
such a low-order statistic only as the by-product of calculating the full
distribution. One would
like to have a more direct variational method for any statistic of interest.

In fact, it is well known in various contexts that such statistical quantities
as expectations, correlations, etc. are
characterized by a minimum principle for a certain functional. In (Euclidean)
field theory this functional is called the
``effective action,'' and was first rigorously investigated by Symanzik in
\cite{Sym}. In nonequilibrium statistical
mechanics the variational principle associated to the effective action was
pointed out some time ago by Graham
\cite{Gra}. The fact that averages of suitable distributions are characterized
by a minimum principle is also standard
in probability theory: see Section 3 of \cite{Str}. Such a principle has a very
general basis and, indeed, its
origin is the same as that of the familiar equilibrium variational principles
of maximum entropy, minimum free-energy, etc.
Closely related ideas have been exploited recently to develop moment-closure
hierarchies for kinetic theories \cite{CDL}.
We shall give here a self-contained discussion of the least-action principle,
following the accounts in \cite{Sym,Str}.

\newpage

The main requirement for its validity is {\em finite exponential moments} of
the statistical distribution.
Let us denote by ${\cal P}$ the probability measure on {\em histories} of our
stochastic dynamics.
Thus, $P_t$ is just the projection (or, marginal) at time $t$ of the
distribution $\cP.$ Then,
what is required is that, integrating over the ensemble of histories
$\{\bx(t):-\infty< t<+\infty\}$,
\be \int \cD\cP(\bx)\,e^{(\bF,\bZ(\bx))}<\infty, \lb{V1} \ee
where $\bF(t)$ is a real-vector valued test function and $(\bF,\bZ)=\int
dt\,\,\bF(t)\bdot\bZ(t).$
If Eq.(\ref{V1}) holds, we may define
\be W[\bF]\equiv \log\left[\int \cD\cP(\bx)\,e^{(\bF,\bZ)}\right], \lb{V2} \ee
which is a cumulant-generating functional of the distribution $\cP.$ It is a
consequence of the
positivity of the distribution and the H\"{o}lder inequality that
\be \int \cD\cP(\bx)\,e^{(\lambda\bF_1+(1-\lambda)\bF_2,\bZ)}\leq
     \left(\int \cD\cP(\bx)\,e^{(\bF_1,\bZ)}\right)^\lambda
     \left(\int \cD\cP(\bx)\,e^{(\bF_2,\bZ)}\right)^{1-\lambda}, \lb{V3} \ee
for $0<\lambda<1,$ or
\be W[\lambda\bF_1+(1-\lambda)\bF_2]\leq \lambda W[\bF_1]+(1-\lambda)W[\bF_2].
\lb{V4} \ee
In other words, $W[\bF]$ is a globally convex functional of its argument.
Observe that this
is a result just of a simple realizability inequality for the distribution
${\cal P}.$
The corresponding conjugate convex functional is
\be \Gamma[\bZ]=\sup_{\bF}((\bF,\bZ)-W[\bF]). \lb{V5} \ee
This is the definition of the {\em effective action} for $\bZ$-histories. Since
$\Gamma[\bZ]$ is also
globally convex under the assumption Eq.(\ref{V1}), it follows that it has an
absolute
minimum (possibly nonunique if $\Gamma$ is not strictly convex). In fact,
\be \Gamma[\bZ]\geq 0\,\,\,\,\& \,\,\,\,\Gamma[\overline{\bZ}]=0, \lb{V6} \ee
where
\be \overline{\bZ}(t)=\int \cD\cP(\bx) \bZ(\bx(t)). \lb{V7} \ee
The positivity of $\Gamma$ follows from the fact that $(\bF,\bZ)-W[\bF]=0$
in Eq.(\ref{V5}) for $\bF=\bz.$ Furthermore, by Jensen's inequality
$\log\left[\int \cD\cP(\bx)
\,e^{(\bF,\bZ)}\right]\geq (\bF,\overline{\bZ}).$ Thus,
$(\bF,\overline{\bZ})-W[\bF]\leq 0$ for all $\bF,$ and so
$\Gamma[\overline{\bZ}]=0.$
That the mean is characterized as the point at which $\Gamma$ achieves its
minimum is just the precise
statement of the {\em principle of least effective action}.

All the derivations we have given for the distribution on histories, $\cP,$
could just as well be given
for the single-time stationary distribution, $P_s.$ However, since the latter
is hard to specify, it
is easier to work with a quantity derived from the effective action introduced
above, which is
commonly referred to as the {\em effective potential}. This is obtained from
the full action by defining, for any
time-independent $\bZ$, the time-extended history $\bZ_T(t)$ by
\be \bZ_T(t)\equiv \left\{ \begin{array}{ll}
                                \bZ & \mbox{if $0<t<T$} \cr
                                \obZ & \mbox{ otherwise}.
                                \end{array}
                        \right. \lb{V21} \ee
Then the ``effective potential'' $V[\bx]$ is defined as the infinite-time limit
\be V[\bZ]=\lim_{T\rightarrow +\infty}{{\Gamma[\bZ_T]}\over{T}}. \lb{V22} \ee
The effective potential is appropriate to determine expected values in the
time-invariant ground
state of the theory $\Omega^R=P_s.$

The effective potential has a direct significance in terms of the statistics of
the {\em empirical time-average}:
\be \overline{\bZ}_T\equiv {{1}\over{T}}\int_0^T dt \,\,\bZ(t). \lb{empavrg}
\ee
For an ergodic process, this random variable converges as $T\rightarrow\infty$
to the ensemble-average, $\overline{\bZ}_T
\rightarrow \overline{\bZ},$ almost surely in every realization. However,
fluctuations away from the expected
behavior should furthermore occur with a small probability, decaying
asymptotically for large $T$ as
\be {\rm Prob}\left(\overline{\bZ}_T\approx \bZ\right)\sim \exp\left(-T\cdot
V[\bZ]\right). \lb{LD} \ee
This is a refinement of the standard ergodic hypothesis. It will hold when the
limit in Eq.(\ref{V22}) exists,
or, equivalently, if the similar limit, $\lim_{T\rightarrow
+\infty}{{1}\over{T}}W[\bh_T]=\lambda[\bh]$, exists.
These are standard results of ``large deviations'' in probability theory
\cite{Ell,Var}. In fact, what is in physics
referred to as the ``effective potential'' coincides for stochastic dynamics
with the (level-1) rate function in
the Donsker-Varadhan large-deviations theory for ergodic Markov processes. The
probabilistic interpretation of the
effective potential seems to have been first pointed out in quantum field
theory by Jona-Lasinio \cite{GJL}.
Such a large-deviations hypothesis as Eq.(\ref{LD}) was conjectured some time
ago by Takahashi for deterministic
dynamical systems with sufficiently chaotic solutions \cite{Tak}, and rigorous
theorems have been proved under suitable
hypotheses (e.g. see \cite{Kif1,Kif2}). In this context the effective potential
is simply related to the Kolmogorov-Sinai
entropy. The earliest origins of the above fluctuation hypothesis in
statistical physics appear in the 1931 ``Onsager
principle'', as discussed by Oono in \cite{Oon}.

It follows from our assumptions that the effective potential is nonnegative,
$V(\bZ)\geq 0$, convex, $\lambda_1V(\bZ_1)+
\lambda_2V(\bZ_2)\geq V(\lambda_1\bZ_1+\lambda_2\bZ_2),
\,\,\,\lambda_1+\lambda_2=1$, and vanishes only at the
ensemble-mean, $V(\obZ)=0$.\footnote{The structure of the effective potential
may be more complex if there is
``ergodicity-breaking'' associated to multiple ergodic measures. In that case,
there may be a convex set of points $\bZ$
with nonempty interior on which $V(\bZ)$ vanishes. This would be the case if a
so-called non-equilibrium phase-transition
occurred. The important applications of the effective potential in quantum
field theory appeared precisely in this
type of situation, where basic symmetries of the quantum Hamiltonian are
spontaneously broken by the occurrence
of multiple ground states. Similiar phenomena may be expected in
infinite-volume nonequilibrium systems, especially
in the parameter range after the first bifurcation from a unique laminar
solution but before transition to fully-developed
turbulence has occurred.} In the next section we develop a practical method for
approximately calculating the effective
potential. Because of the connection of the effective potential with
fluctuations of the empirical mean, Eq.(\ref{LD}),
it is very unlikely that a closure approximation which violates the basic
positivity and convexity properties of the
effective potential can yield a reasonable result for the ensemble-average
itself.

\noindent {\em (ii) Variational Characterization of Effective Potential}

We now show how the effective potential $V(\bZ)$ is related to the Hamiltonian
$\cH[\Psi^R,\Psi^L]$ discussed
before by means of a {\em constrained variation}. A similar result was proved
by Symanzik in Euclidean field
theory \cite{Sym}. In our case, a modification is required associated to the
non-self-adjoint character of $\hL.$
More precisely, we have
\begin{Th}
The effective potential
\be V[\bZ]=\lim_{T\rightarrow +\infty}{{1}\over{T}}\Gamma[\bZ_T], \lb{V58} \ee
for a stationary Markov process is the value at the extremum point of the
functional
\be V[\Psi^R,\Psi^L]= -\cH[\Psi^R,\Psi^L]. \lb{59a} \ee
varying over all pairs of state vectors $\Psi^R,\Psi^L$ subject to the
constraints
\be \langle\Psi^L,\Psi^R\rangle=1 \lb{V60} \ee
and
\be \langle\Psi^L,\hBZ\cdot\Psi^R\rangle=\bZ. \lb{V61} \ee
\end{Th}
Here $\hBZ$ is the operator of multiplication by $\bZ(\bx).$  Although the
original version of the theorem
required just one trial state, there now must be {\em two independent trial
states.}

Nevertheless, the proof is similar to the original one of Symanzik \cite{Sym}.
Let $\Omega^R=P_s,\Omega^L\equiv 1.$
Then the generating functional $W[\bh]$ introduced above may be represented in
the operator formulation by
\be W[\bh]=\log\langle\Omega^L, {\rm T}\exp\left(\int_0^T
dt\,\hL_\bh(t)\right)\cdot\Omega^R\rangle, \lb{V62} \ee
where ${\rm T}$ denotes time-ordering (increasing right to left) and
\be \hL_\bh(t)=\hL+\bh(t)\bdot\hBZ . \lb{V63} \ee
No time-dependence is required for the coordinate operators because the
exponential factors
automatically introduce the correct Heisenberg picture operators after
differentiating and setting $\bh$ to zero.
We note then that for a {\em static} field $\bh$ in the limit $T\rightarrow
+\infty,$
\begin{eqnarray}
\exp(W[\bh_T]) & =       &  \langle\Omega^L,\exp\left(T\cdot \hL_\bh
\right)\cdot\Omega^R\rangle \cr
             \,& \approx &
\langle\Omega^L,\Omega^R[\bh]\rangle\langle\Omega^L[\bh],\Omega^R\rangle
                                    \times \exp(T\cdot\lambda[\bh]), \lb{V64}
\end{eqnarray}
where $\lambda[\bh]$ is the eigenvalue of the ``perturbed operator''
\be \hL_\bh=\hL+\bh\bdot\hBZ \lb{V65} \ee
with the {\em largest real part} and $\Omega^R[\bh],\Omega^L[\bh]$ are the
associated right and left
``ground state'' eigenvectors:
\be \hL_\bh|\Omega^R[\bh]\rangle=\lambda[\bh]|\Omega^R[\bh]\rangle, \lb{V65a}
\ee
and
\be \hL_\bh^\dagger|\Omega^L[\bh]\rangle=\lambda^*[\bh]|\Omega^L[\bh]\rangle.
\lb{V66} \ee
Furthermore, we can see that
\be {{\partial W[\bh_T]}\over{\partial h_n}}= T\cdot z_n[\bh]+o(T), \lb{V66a}
\ee
with
\be z_n[\bh]= \langle\Omega^L[\bh],\HZ_n\cdot\Omega^R[\bh]\rangle. \lb{V67} \ee
This can be obtained from the formula
\begin{eqnarray}
\exp(W[\bh_T]){{\partial W[\bh_T]}\over{\partial h_n}}
               & = & \langle\Omega^L,{{\partial}\over{\partial
h_n}}\exp\left(T\cdot \hL_\bh \right)\cdot\Omega^R\rangle \cr
             \,& = & \langle\Omega^L,
\Omega^R[\bh]\rangle\langle\Omega^L[\bh],\Omega^R\rangle
      \langle\Omega^L[\bh], {{\partial}\over{\partial h_n}}\exp\left(T\cdot
\hL_\bh \right)\cdot\Omega^R[\bh]\rangle \cr
             \,&   &
\,\,\,\,\,\,\,\,\,\,\,\,\,\,\,\,\,\,\,\,\,\,\,\,\,\,\,\,\,\,\,\,\,\,\,\,
                          +O\left(e^{-T\cdot\Delta\lambda}\right), \lb{V67a}
\end{eqnarray}
where $\Delta\lambda$ is the spectral gap between the real parts of the
``ground state'' eigenvalue and
the next highest eigenvalue. We have used the well-known fact that, for any
one-parameter family
of operators $\hL(h)$ depending smoothly on a parameter $h,$
\be {{\partial}\over{\partial h}}\exp(\hL(h))=\exp(\hL(h))\varphi(-{\rm
Ad}\hL(h))
                     \left[{{\partial \hL(h)}\over{\partial h}}\right],
\lb{V68} \ee
where ${\rm Ad}\hL$ denotes the ``adjoint operator'' defined by the commutator,
\be ({\rm Ad}\hL)[\hat{O}]=[\hL,\hat{O}], \lb{V69} \ee
and $\varphi(z)$ is the entire function
$\varphi(z)=(e^z-1)/z=1+{{1}\over{2!}}z+{{1}\over{3!}}z^2\cdots.$
See \cite{HS}. Since
\be \langle\Omega^L[\bh], [\hL_\bh,\hat{O}]\cdot\Omega^R[\bh]\rangle=0,
\lb{V71} \ee
for any operator $\hat{O},$ only the first term survives in the expansion of
$\varphi$
when substituted into the first term of formula Eq.(\ref{V67a}). This yields
Eq.(\ref{V66a}).

Now let us consider the variational problem. If we incorporate the constraints
by suitable
Lagrange multipliers, then the variational equation is just
\be \delta\left[ -\langle\Psi^L, \hL\cdot\Psi^R\rangle-\bh\bdot\langle\Psi^L,
\hBZ\cdot\Psi^R\rangle
                +\lambda\langle\Psi^L, \Psi^R\rangle\right]=0, \lb{V72} \ee
or
\be \langle\delta\Psi^L, \left(\hL_\bh-\lambda\right)\Psi^R\rangle
       +\langle\Psi^L, \left(\hL_\bh-\lambda\right)\delta\Psi^R\rangle=0.
\lb{V72c} \ee
In other words, there are infinitely many stationary points of the functional
$V[\Psi^R,\Psi^L]$ subject
to the constraints. They consist precisely of pairs
$(\Psi_\nu^R[\bh],\Psi_\nu^L[\bh])$ of
eigenvectors of $\hL_\bh$,
\be  \hL_\bh|\Psi_\nu^R[\bh]\rangle=\lambda_\nu[\bh]|\Psi^R_\nu[\bh]\rangle,
\lb{V72a} \ee
and
\be \hL_\bh^\dagger|\Psi_\nu^L[\bh]\rangle=
          \lambda^*_\nu[\bh]|\Psi_\nu^L[\bh]\rangle, \lb{V72b} \ee
corresponding to different branches of eigenvalues
$\lambda_\nu[\bh],\nu=0,1,2,...$ To be precise, we should
consider the stationary point corresponding to the branch with largest real
part for each $\bh,$ that is, the pair of
``ground state'' eigenvectors $(\Omega^R[\bh],\Omega^L[\bh])$ introduced above.
For small enough $\bh$ this corresponds
to the eigenvalue branch with $\lambda(\bz)=0,$ because the spectrum of $\hL$
is all in the left half of the complex
$\lambda$-plane, ${\rm Re}\lambda<0,$ except for a simple eigenvalue at
$\lambda=0.$ See \cite{GS} and \cite{Kras}.
We refer to this as the ``zero-branch'' of eigenvalues.

Applying then the left eigenvector to the eigen-equation of the right vector
and using the constraints gives
\be  \langle\Omega^L[\bh], \hL\cdot\Omega^R[\bh]\rangle+
\bh\bdot\bZ[\bh]=\lambda[\bh], \lb{V73} \ee
and thus
\begin{eqnarray}
-\langle\Omega^L[\bh], \hL\cdot\Omega^R[\bh]\rangle & = &
\bh\bdot\bZ[\bh]-\lambda[\bh] \cr
    \, & = & {{1}\over{T}}\left[ \langle\bh_T,{{\delta
W}\over{\delta\bh}}[\bh_T]\rangle-W[\bh_T]\right]+o(1), \cr
    \, & = & {{1}\over{T}}\Gamma[\bZ_T]+o(1). \lb{V74}
\end{eqnarray}
The first quantity is independent of $T,$ so that we see taking the limit
$T\rightarrow +\infty$ that
\be -\langle\Omega^L[\bh], \hL\cdot\Omega^R[\bh]\rangle = V[\bZ], \lb{V75} \ee
as was claimed. $\,\,\,\,\Box$

We have given only a formal proof of the theorem without a careful statement of
the conditions, which would certainly
involve spectral properties of the ``Liouville operator'' $\hL,$ etc. The
assumption of a spectral gap may be
stronger than required. The above variational characterization of the effective
potential is, in fact, equivalent to a
spectral characterization of the potential which has been rigorously
established in the Donsker-Varadhan theory
\cite{Var,Ell,Str}. In that case it is shown, under suitable conditions, that
$V[\bZ]=\sup_{\bh}\left(\bZ\bdot\bh
-\lambda[\bh]\right)$ where $\lambda[\bh]$ is the ``principal eigenvalue'' of
the operator $\hL_\bh=\hL+\bh\bdot\hBZ.$
The equivalence of these two characterizations follows from the preceding
formal proof. The representation of the
potential $V[\bZ]$ as a Legendre transform of $\lambda[\bh]$ is entirely
analogous to the representation of the
entropy in equilibrium lattice spin systems as the Legendre transform of the
free-energy, where the latter
is determined as the leading eigenvalue of the transfer matrix. For
deterministic dynamics the existence of a spectral
gap in the so-called ``Perron-Frobenius operator'' has been established only
for a few special cases, such as the work
of Pollicot and Ruelle on Axiom A systems \cite{Rue1}. The eigenvalue
$\lambda[\bh]$ in that context is a particular
case of the topological pressure $P(\varphi)$: see \cite{Kif2} (or \cite{BS}
for an introduction). For example,
in the work of Ruelle \cite{Rue3} on expanding maps $f$ of compact spaces $X$,
the effective potential
would coincide with $P(\varphi)$ for the choice
$\varphi(\bx)=-\ln|f'(\bx)|+\bh\bdot\bZ(\bx)$. Here $|f'(\bx)|$
is the Jacobian determinant of the map, and its logarithm, $\ln|f'(\bx)|$, is
the ``Hamiltonian'' in the thermodynamic
formalism for expanding maps.

\newpage

\noindent {\em (iii) Rayleigh-Ritz Approximation of the Effective Potential}

We outline a simple variational method of Rayleigh-Ritz type to approximate the
effective potential and, thereby, the
ensemble means. The ansatz used previously for $\Psi^R,\Psi^L$ may need to be
replaced by ``augmented ansatz'' $\oPsi^R,
\oPsi^L$.  The reason is that the left ground state, under the imposed
constraint, is no longer $1$ identically
and the constant component must be allowed to vary. In other words, we must
augment the linear ansatz Eq.(\ref{ansl*})
for the left ground state, by setting
\be \oPsi^L(\balpha,\balpha^L)=\sum_{i=0}^N\alpha_i^L\psi_i^L(\balpha),
\lb{nw0} \ee
Here the test function
\be \psi_0^L(\bx;\balpha)\equiv 1\lb{nw00} \ee
is included with an adjustable parameter $\alpha^L_0.$ Of course, with the
orthogonal expansion ansatz
Eqs.(\ref{oprans}),(\ref{oplans}), the constant term (zero-degree polynomial)
is already included. However, if
it was not originally, it should now be added, and an additional free parameter
$\alpha_0$ should be added to the
PDF ansatz $P=\Psi^R(\balpha)$ as well. The most natural way to do so is to
simply replace the normalized density
$\Psi^R\geq 0$ by
\be \oPsi^R(\bx;\obalpha)=\alpha_0\Psi^R(\bx;\balpha), \lb{augRans} \ee
where $\alpha_0$ denotes an arbitrary normalization factor:
\be \int d\bx\,\,\oPsi^R(\bx;\obalpha)=\alpha_0. \lb{normfac} \ee
Because $\oPsi^L\neq 1$ under the constraint, unit normalization of $\oPsi^R$
is no longer required, but,
instead, the overlap condition $\langle\oPsi^L,\oPsi^R\rangle=1$ must be
maintained. Notice that we use the notations
$\obalpha,\obalpha^L$ simply to indicate the parameter vectors
$\balpha,\balpha^L$ along with the additional
zero-components $\alpha_0,\alpha^L_0.$ We shall refer to the new ansatz
Eqs.(\ref{augRans}),(\ref{nw0}) as the
{\em natural augmentation}. While others can be contrived, this is the simplest
extended ansatz and likely to be
the most generally useful. \footnote{Despite this, some of our arguments below
do not apply to the natural augmentation!
We will point out where this occurs in the later discussion. This is really a
technical issue, since all of the
{\em results} discussed hereafter still hold for the natural augmentation and
it is only the proofs which need to be
changed somewhat. Rather than complicate the discussion, we have decided to
present proofs under the simplest assumptions.
These are satisfied, for example, by the orthogonal expansion ansatz. The
natural augmentation is discussed in detail
elsewhere \cite{AlEy3}.} Note it is not necessary to have a closed-form
expression for $\Psi^R$, but it is enough only
to be able to calculate averages such as
\be \om_i(\obalpha)=\langle\psi_i^L(\obalpha)\rangle_{\obalpha}. \lb{augmom}
\ee
and
\be \oV_i(\obalpha,\bh)=\langle\hat{L}_\bh^\dagger
          \psi_i^L(\obalpha)\rangle_{\obalpha}, \lb{augvec} \ee
with $i=0,1,...,N.$ In the most practical PDF closures, the ansatz
$\Psi^R(\bx;\balpha)$ will be given, not explicitly,
but instead by averages with respect to ``surrogate'' random variables
$\bX_\balpha$ whose distributions
are parametrized by $\balpha.$ From the joint ansatz for $\oPsi^H,\,H=L,R,$ an
approximation to the effective
potential is then obtained:
\be V_*(\bZ)= -\langle\oPsi_*^L,\hat{L}\oPsi_*^R\rangle, \lb{nw1} \ee
where $\oPsi^L_*=\oPsi^L(\obalpha_*(\bh),\obalpha_*^L(\bh))$ \&
$\oPsi^R_*=\oPsi^R(\obalpha_*(\bh)),$ and
the parameters $\obalpha_*^L(\bh),\obalpha_*(\bh),$ and $\bh=\bh_*(\bZ)$ are to
be determined as follows.

Incorporating as before the constraints by
suitable Lagrange multipliers $\lambda$ and $\bh,$ the extremum point within
the ansatz is obtained by
varying the function
\be F(\obalpha,\obalpha^L)\equiv
-\langle\oPsi^L(\obalpha,\obalpha^L),\hL_\bh\oPsi^R(\obalpha)\rangle
+\lambda\langle\oPsi^L(\obalpha,\obalpha^L),\oPsi^R(\obalpha)\rangle, \lb{nw1a}
\ee
of the parameters $\obalpha,\obalpha^L.$ First, by variation of the
$\obalpha$-parameters, one obtains the equation
\be \bA(\obalpha,\bh)\bdot\obalpha^L= \lambda\bB(\obalpha)\bdot\obalpha^L
\lb{nw2} \ee
with the matrices $\bA(\obalpha,\bh)$ and $\bB(\obalpha)$ defined by
\be
A_{ij}(\obalpha,\bh)={{\partial}\over{\partial\oalpha_i}}\oV_j(\obalpha,\bh)
\lb{nw3} \ee
and
\be B_{ij}(\obalpha)={{\partial}\over{\partial\oalpha_i}}\om_j(\obalpha)
\lb{nw3a} \ee
for $i,j=0,1,...,N.$ Eq.(\ref{nw2}) has the form of a {\em generalized
eigenvalue problem} \cite{Wk,Kat}.
The parameter vector $\obalpha^L(\obalpha,\bh)$ is to be determined as the
generalized eigenvector associated to the
``leading'' eigenvalue.

However, the proper definition of this last quantity requires some discussion.
In the original infinite-dimensional setting, the ``leading'' eigenvalue was
defined to be that with largest real
part and, for $\bh$ small enough, it coincides with the ``zero-branch'' passing
through 0 for $\bh=\bz.$ On the other hand,
within an approximation such as we consider here, these two quantities need no
longer coincide, although both exist.
An eigenvalue branch $\lambda(\obalpha,\bh)$ such that
$\lambda(\obalpha,\bz)=0$ exists always with the associated
eigenvector $\oalpha^L_i=\delta_{i0}$ at $\bh=\bz.$ Likewise, an eigenvalue
with a real part---
denoted $\Lambda(\obalpha,\bh)$---of largest value will certainly exist.
Because the two quantities $\lambda(\obalpha,\bh)$
and $\Lambda(\obalpha,\bh)$ are possibly distinct, either may be plausibly used
as the basis of an approximate calculation.
However, there are compelling reasons to prefer the use of
$\lambda(\obalpha,\bh).$ Most importantly, it is only due to
$\lambda(\obalpha,\bz)=0$ that $\obalpha_*(\bz)=\obalpha_*$ coincides with one
of the fixed points of the $\bh=\bz$
vector field $\obV(\obalpha)$ (see below). Also, as a practical matter, it will
generally be easier to compute
$\lambda(\obalpha,\bh)$ than $\Lambda(\obalpha,\bh),$ whose calculation
requires a determination of the entire spectrum
of $\bA(\obalpha,\bh).$ Actually, all of these considerations are rather
academic. If $\Lambda(\obalpha,\bh)>
\lambda(\obalpha,\bh)=0$ at $\bh=\bz$, then the stability matrix ${{\partial
\obV}\over{\partial\obalpha}}(\obalpha)=
[\bA(\obalpha,\bz)]^\top$ has an eigenvalue with positive real part. If this
were to occur at the starting point
$\obalpha_*$, that point would be linearly unstable under the dynamical flow of
the vector field $\obV(\obalpha)$.
That alone would be enough to disqualify the point $\obalpha_*$ from physical
interest. On the other hand, if
$\Lambda(\obalpha,\bh)=\lambda(\obalpha,\bh)$ at $\bh=\bz$, then, except for
degenerate cases, this will also
be true in a small interval of $\bh$ about $\bz$ and no distinction need be
made. It will be explained below that
the approximate potential $V_*(\bZ)$ calculated from $\lambda(\obalpha,\bh)$
necessarily has the approximate mean
\be \obZ_*\equiv \int d\bx\,\,\bZ(\bx)\cdot\Omega^R_*(\bx), \lb{approxmn} \ee
as a critical point, with $V_*(\obZ_*)=0,$ but that $V_*(\bZ)$ need no longer
be convex at $\obZ_*.$

Returning, then, to the specification of the approximation scheme, we next
determine $\obalpha_*(\bh)$
as the value of $\obalpha$ satisfying the variational equation under the
parameters $\obalpha^L$:
\be \oV_i(\obalpha,\bh)=\lambda(\obalpha,\bh)\om_i(\obalpha), \lb{nw6} \ee
$i=0,1,...,N.$ This may be thought of as a type of ``nonlinear eigenvalue
condition'' and $\obalpha_*(\bh)$ as the
associated eigenvector. Since $\lambda(\obalpha,\bz)=0,$ it is a consequence of
this definition that
\be \obalpha_*(\bz)=\obalpha_*, \lb{FPcond} \ee
with $\obalpha_*$ a fixed point of the dynamical vector $\obV(\obalpha)$
defined in Eq.(\ref{dynvec}).
As long as the stability matrix ${{\partial
\obV}\over{\partial\obalpha}}(\obalpha_*)$ is non-singular,
the implicit function theorem guarantees that Eq.(\ref{nw6}) has a solution for
at least some small
interval of $\bh$ about $\bz$. \footnote{This is the property which is not
satisfied by the ``natural
augmentation.'' In fact, it is not hard to show that with that choice
\be {{\partial
\obV}\over{\partial\oalpha}}(\oalpha_*)=\left(\begin{array}{cc}
                                                               0 & \bz \cr
                                                               \bz & {{\partial
\bV}\over{\partial\alpha}}(\alpha_*) \cr
                                                               \end{array}
                                                         \right). \lb{stabmat}
\ee
Clearly, this matrix is singular. However, as we have already noted, it is only
the present proofs which fail and the
results themselves, proved here assuming non-singularity, still hold for the
``natural augmentation'' \cite{AlEy3}.}
For practical computation, a Newton-Raphson or other root-finding algorithm
may be employed (see \cite{numrec}, Ch.9), starting with $\obalpha_*$ at
$\bh=\bz$ and tracking a sequence
of roots $\obalpha_*(\bh_k)$ iteratively for $\bh_k$ of increasing magnitude.
If the starting
ansatz $\oPsi^R,\oPsi^L$ has more than one acceptable fixed point, then any of
them may be used as a basis for the
calculation. Next, $\obalpha_*^L(\bh)$ is defined as
$\obalpha^L(\obalpha_*(\bh),\bh)$ with its normalization
fixed by the constraint $\langle\oPsi_*^L,\oPsi^R_*\rangle=1.$ This allows one
to define the function
\be \bZ_*(\bh)\equiv
\langle\oPsi^L\left(\obalpha_*^L(\bh),\obalpha_*(\bh)\right),
\,\hat{\BZ}\cdot\oPsi^R\left(\obalpha_*(\bh)\right)\rangle, \lb{nw7} \ee
and to determine $\bh$ thereby as the value $\bh_*(\bZ)$ of its inverse
function at $\bx.$ It should be remarked that
both $\obalpha_*(\bh)$ and $\obalpha_*^L(\bh)$ are real vectors, at least for
small enough $\bh,$ and therefore
$\bZ_*[\bh]$ is a real-vector too. The eigenvalue $\lambda(\obalpha,\bh)$ will
be real for $\bh$ sufficiently near
$\bz$ and, in that case, the associated generalized eigenvector
$\obalpha^L(\obalpha,\bh)$ for the real matrices
$\bA(\obalpha,\bh),\bB(\obalpha)$ will also be real. We observe for $\bh=\bz$
that $\bZ_*[\bz]=\obZ_*.$

These prescriptions complete our recipe for the Rayleigh-Ritz approximation to
the effective potential $V(\bZ).$
We now establish an important representation for $V_*(\bZ).$ Let us define
\be \lambda_*(\bh)\equiv \lambda(\obalpha_*(\bh),\bh), \lb{geneig} \ee
in terms of the quantities introduced above. We now prove
\begin{Prop}
The approximate effective potential $V_*(\bZ)$ is a formal Legendre transform
of $\lambda_*(\bh)$; that is,
\be {{\partial\lambda_*}\over{\partial\bh}}(\bh)=\bZ_*(\bh) \lb{difrel} \ee
and
\be V_*(\bZ)=\bZ_*(\bh)\bdot\bh-\lambda_*(\bh), \lb{formLeg} \ee
for $\bh=\bh_*(\bZ).$
\end{Prop}

\noindent {\em Proof}: Setting
\be
\oPsi^L_*(\bx;\bh)=\sum_{i=0}^N\oalpha_{*i}^L(\bh)\psi^L_i(\obalpha_*(\bh)),
\lb{Lgrst} \ee
and
\be \oPsi^R_*(\bx;\bh)=\oPsi(\bx;\obalpha_*(\bh)), \lb{Rgrst} \ee
we observe the overlap condition
$\langle\oPsi^L_*(\bh),\oPsi^R_*(\bh)\rangle=1$ becomes simply
\be \sum_{i=0}^N\oalpha_{*i}^L(\bh)\om_i(\obalpha_*(\bh))=1. \lb{ovlap} \ee
We next show that
\be \langle\oPsi^L_*(\bh),\hL_\bh\cdot\oPsi^R_*(\bh)\rangle=\lambda_*(\bh).
\lb{eigcond} \ee
In fact,
\begin{eqnarray}
\langle\oPsi^L_*(\bh),\hL_\bh\cdot\oPsi^R_*(\bh)\rangle
                         & = &
\sum_{i=0}^N\oalpha_{*i}^L(\bh)\oV_i(\obalpha_*(\bh),\bh) \cr
                       \,& = &
\lambda_*(\bh)\sum_{i=0}^N\oalpha_{*i}^L(\bh)\om_i(\obalpha_*(\bh)) \cr
                       \,& = & \lambda_*(\bh), \lb{verif}
\end{eqnarray}
where the first line follows using the linear ansatz, Eq.(\ref{Lgrst}) above,
the second line follows from
the ``nonlinear eigenvalue condition'' Eq.(\ref{nw6}), and the last line
follows from the overlap condition
Eq.(\ref{ovlap}). Now it is easy to see that
\begin{eqnarray}
V_*(\bZ) & = & -\langle\oPsi^L_*(\bh),\hL\cdot\oPsi^R_*(\bh)\rangle \cr
       \,& = & \langle\oPsi^L_*(\bh),\hBZ\cdot\oPsi^R_*(\bh)\rangle\bdot\bh-
\langle\oPsi^L_*(\bh),\hL_\bh\cdot\oPsi^R_*(\bh)\rangle \cr
       \,& = & \bZ_*(\bh)\bdot\bh-\lambda_*(\bh), \lb{Legeq}
\end{eqnarray}
which is Eq.(\ref{formLeg}).

The verification of Eq.(\ref{difrel}) is a straightforward but somewhat tedious
calculation.
Using once more the basic expression Eq.(\ref{eigcond}) for $\lambda_*(\bh),$
one finds by
differentiation that
\be {{\partial\lambda_*}\over{\partial\bh}}(\bh)=\bZ_*(\bh)
+\langle{{\partial\oPsi^L_*}\over{\partial\bh}}(\bh),
                               \hL_\bh\cdot\oPsi^R_*(\bh)\rangle
+\langle\oPsi^L_*(\bh),\hL_\bh\cdot{{\partial\oPsi^R_*}
               \over{\partial\bh}}(\bh)\rangle. \lb{deriveig} \ee
Furthermore, calculation yields for the second term
\begin{eqnarray}
\,& & \langle{{\partial\oPsi^L_*}\over{\partial\bh}}(\bh),
                                  \hL_\bh\cdot\oPsi^R_*(\bh)\rangle=
\sum_{i=0}^N\left({{\partial}\over{\partial\bh}}\oalpha_{*i}^L(\bh)
                          \right)\lambda_*(\bh)\om_i(\obalpha_*(\bh)) \cr
\,& &\,\,\,\,\,\,\,\,\,\,\,\,\,\,\,\,\,\,\,\,\,\,\,\,\,\,\,\,
\,\,\,\,\,\,\,\,\,\,\,\,\,\,+\sum_{i,j=0}^N\oalpha_{*i}^L(\bh)
\langle\hL^\dagger_\bh{{\partial\psi_i^L}\over{\partial\oalpha_j}}
 (\obalpha_*(\bh))\rangle_{\obalpha_*(\bh)}
              {{\partial\oalpha_{*j}}\over{\partial\bh}}(\bh), 
\lb{firstlong}
\end{eqnarray}
where the ``nonlinear eigenvalue condition'' Eq.(\ref{nw6}) was used in the
first sum on the righthand side. Likewise,
for the third term in Eq.(\ref{deriveig})
\begin{eqnarray}
\,& & \langle\oPsi^L_*(\bh),\hL_\bh\cdot{{\partial\oPsi^R_*}
                             \over{\partial\bh}}(\bh)\rangle=
\sum_{i=0}^N\oalpha_{*i}^L(\bh)\cdot\lambda_*(\bh)\left({{\partial}
                    \over{\partial\bh}}\om_i(\obalpha_*(\bh))\right) \cr
\,& &\,\,\,\,\,\,\,\,\,\,\,\,\,\,\,\,\,\,\,\,\,\,\,\,\,\,\,\,
     \,\,\,\,\,\,\,\,\,\,\,\,\,\,-\sum_{i,j=0}^N\oalpha_{*i}^L(\bh)
\langle\hL^\dagger_\bh{{\partial\psi_i^L}\over{\partial\oalpha_j}}
          (\obalpha_*(\bh))\rangle_{\obalpha_*(\bh)}
    {{\partial\oalpha_{*j}}\over{\partial\bh}}(\bh), \lb{secondlong}
\end{eqnarray}
where the generalized eigenvalue equation Eq.(\ref{nw2}) was used in the first
sum on the righthand side . Adding the two
contributions, the last terms of each cancel and the result is
\begin{eqnarray}
\,&  & \langle{{\partial\oPsi^L_*}\over{\partial\bh}}(\bh),
                               \hL_\bh\cdot\oPsi^R_*(\bh)\rangle
+\langle\oPsi^L_*(\bh),\hL_\bh\cdot{{\partial\oPsi^R_*}
                                   \over{\partial\bh}}(\bh)\rangle \cr
\,&  & \,\,\,\,\,\,\,\,\,\,\,\,\,\,\,\,\,\,\,\,\,
=\sum_{i=0}^N\left[\left({{\partial}\over{\partial\bh}}
             \oalpha_{*i}^L(\bh)\right)\lambda_*(\bh)\om_i(\obalpha_*(\bh))
+\oalpha_{*i}^L(\bh)\cdot\lambda_*(\bh)\left({{\partial}\over{\partial\bh}}
                              \om_i(\obalpha_*(\bh))\right)\right] \cr
\,&  & \,\,\,\,\,\,\,\,\,\,\,\,\,\,\,\,\,\,\,\,\,
=\lambda_*(\bh){{\partial}\over{\partial\bh}}\left[\sum_{i=0}^N
                \oalpha_{*i}^L(\bh)\om_i(\obalpha_*(\bh))\right] \cr
\,&  & \,\,\,\,\,\,\,\,\,\,\,\,\,\,\,\,\,\,\,\,\,
      =0. \lb{zerover}
\end{eqnarray}
The constant overlap, Eq.(\ref{ovlap}), was invoked in the last line. Thus,
${{\partial\lambda_*}\over{\partial\bh}}(\bh)=\bZ_*(\bh).$ It may be worth
remarking that this result is
a nonlinear generalization of the Hellmann-Feynman theorem used in
quantum-mechanical perturbation theory. $\,\,\,\,\Box$

\vspace{.1in}

\noindent It is a consequence of this proposition that
\be V_*(\obZ_*)=0 \,\,\,\,\&\,\,\,\, {{\partial
V_*}\over{\partial\bZ}}(\obZ_*)=\bz. \lb{conseq} \ee
Indeed, since $\bZ_*(\bz)=\obZ_*$ and $\lambda_*(\bz)=0,$ the first follows
directly from Eq.(\ref{formLeg}).
For the second, we use the simple result of Eq.(\ref{formLeg}) that
\be {{\partial V_*}\over{\partial\bZ}}(\bZ)=\bh_*(\bZ) \lb{derivV} \ee
and $\bh_*(\obZ_*)=\bz.$ Hence we conclude that the properties
Eq.(\ref{conseq}), which hold for the
{\em exact} effective potential, are automatically guaranteed to hold in the
Rayleigh-Ritz approximation.
However, the important property of {\em convexity} of $V_*(\bZ)$ is not
guaranteed. All that can be inferred
from Eq.(\ref{formLeg}) is that $V_*(\bZ)$ is convex in $\bZ$ if and only if
$\lambda_*(\bh)$ is convex in $\bh.$

Let us first, however, note a useful simplification. Just as was discussed in
Section 2.ii, it is
very convenient here also to replace the parameters $\obalpha$ by the moments
$\obm.$ Assuming that the
matrix $\bB(\obalpha)={{\partial\obm}\over{\partial\obalpha}}$ defined in
Eq.(\ref{nw3a}) is nonsingular,
then the relation $\obm=\obm(\obalpha)$ may be inverted, at least locally, to
give $\obalpha(\obm)$
as a function of $\obm.$ Therefore, the $\obm$ may be used as parameters
instead of the $\obalpha,$ writing
as well $\psi^L(\obm)=\psi^L(\balpha(\obm)),
\oPsi^R(\obm)=\oPsi^R(\obalpha(\obm))$ without any possibility of confusion.
In this case, the equation obtained under variation of the $\obm$-parameters
reduces to an ordinary eigenvalue problem:
\be \bA(\obm,\bh)\bdot\obalpha^L= \lambda\bdot\obalpha^L \lb{nw10} \ee
with the matrix $\bA(\obm,\bh)$ defined similarly as before:
\be A_{ij}(\obm,\bh)\equiv{{\partial}\over{\partial\om_i}}\oV_j(\obm,\bh)
\lb{nw11} \ee
and
\be \oV_i(\obm,\bh)\equiv \langle\hat{L}_\bh^\dagger\psi_i^L(\obm)\rangle_\obm,
\lb{nw12} \ee
Once again, $\lambda(\obm,\bh)$ may be taken as the ``leading'' eigenvalue and
$\obalpha^L(\obm,\bh)$ its associated
eigenvector. Likewise, an equation may be obtained for $\obm_*(\bh)$ by varying
$\obalpha^L,$ which is now simply
\be \obV(\obm,\bh)=\lambda(\obm,\bh)\obm. \lb{nw13} \ee
With these additional simplifications, the procedure to calculate $V_*(\bZ)$ is
otherwise the same as before.

In calculating the approximation $V_*(\bZ)$ by the Rayleigh-Ritz method, one
obtains as well approximations to
$\Omega^H,\,H=L,R.$ Since it requires more work to impose the constraints, it
may seem that nothing has been gained
and, even, something has been lost. However, a moderately good ansatz
$\Psi^H(\balpha,\balpha^H)$ may yield rather
poor results for $\Omega^R$ and yet quite good results for $\obZ.$ It is useful
to calculate the effective potential
from the ansatz as a diagnostic since the qualitative features should be
reproduced that $V_*(\bZ)\geq 0$ and that
$\obZ_*$ is a minimum point of $V_*$ with $V_*(\overline{\bZ}_*)=0.$ If one's
only interest
is in the mean values, then these are more realistic criteria of
``acceptability'' of the approximation than to insist,
e.g., that $\Psi_*^R\geq 0$ everywhere. Negative density in an insignificant
region of $\bx$-space might have
very little effect on the approximate average $\obZ_*,$ which could be quite
close to the true average $\obZ.$ On the
other hand, a failure of convexity of $V_*(\bZ)$ would doubtless indicate
serious errors
in $\obZ_*$ as an approximation to $\obZ.$ Such a ``prediction'' would need to
be discarded as spurious.
The condition of convexity of the effective potential is not contained in any
property of the closure dynamics
and it incorporates important additional information from the exact Liouville
dynamics.

\newpage

\noindent {\em (iv) Variational Characterization of Effective Action}

We now show that the time-dependent effective action can also be obtained by a
constrained variation of
the nonequilibrium action functional $\Gamma[\Psi^R,\Psi^L].$  The proof of
this theorem is almost the same as the
proof of a corresponding result in quantum field theory due to Jackiw and
Kerman \cite{JK}. Just as the Symanzik
theorem is a constrained version of the familiar quantum variational principle
for energy eigenvalues and eigenvectors,
the Jackiw-Kerman theorem can be seen as a constrained version of Dirac's
\cite{Dir} variational formulation of the
Schr\"{o}dinger equation (a quantum analogue of Hamilton's principle). In
addition to providing a basis for time-dependent
Rayleigh-Ritz calculations, the Jackiw-Kerman-type theorem establishes the
existence of a Lagrangian functional for
the effective action.
\begin{Th}
The effective action $\Gamma[\bZ]$ for the initial-value problem is the value
at the extremum point of the functional
\be \Gamma[\Psi^R,\Psi^L]= \int^\infty_0
dt\,\,\langle\Psi^L(t),(\partial_t-\hL)\Psi^R(t)\rangle,
\lb{V25} \ee
when that is independently varied over all pairs of time-dependent state
vectors subject to the constraints
for each time $t$:
\be  \langle\Psi^L(t),\Psi^R(t)\rangle=1 \lb{V26} \ee
and
\be  \langle\Psi^L(t),\hBZ\Psi^R(t)\rangle=\bZ(t), \lb{V27} \ee
and also to the boundary conditions
\be |\Psi^R(0)\rangle= P_0\,\,\,\,\&\,\,\,\,\, |\Psi^L(\infty)\rangle\equiv 1.
\lb{76} \ee
\end{Th}
The proof is as follows:

As in the static case, we use the representation
\be W[\bh]=\log\langle\Omega^L,\,{\rm T}\exp\left(\int_0^{\infty}
dt\,\hL_\bh(t)\right)\cdot\Omega^R\rangle, \lb{V62*} \ee
where $\hL_\bh(t)=\hL+\bh(t)\bdot\hBZ$ as before but now
$\Omega^R=P_0,\Omega^L\equiv 1.$ In other words,
\be W[\bh]=\log\langle\Omega^L(t),\Omega^R(t)\rangle, \lb{F1} \ee
where
\be |\Omega^R(t)\rangle={\rm T}\exp\left(\int_0^t
ds\,\,\hL_\bh(s)\right)|\Omega^R\rangle, \lb{F2} \ee
and, if $\overline{{\rm T}}$ denotes ``anti-time-ordering,''
\be |\Omega^L(t)\rangle=\overline{{\rm T}}\exp
                      \left(\int_t^\infty
ds\,\,\hL_\bh^\dagger(s)\right)|\Omega^L\rangle. \lb{F3} \ee
These trajectories are the solutions, respectively, of the initial-value
problem
\be \partial_t|\Omega^R(t)\rangle=
\hL_\bh(t)|\Omega^R(t)\rangle\,\,\,\,\,\,\,\,\,\,\,\,\Omega^R(0)=P_0, \lb{F4}
\ee
and of the final-value problem
\be \partial_t|\Omega^L(t)\rangle=
-\hL_\bh^\dagger(t)|\Omega^L(t)\rangle\,\,\,\,\,\,\,\,\,\,\,\,
                                    \Omega^L(\infty)\equiv 1. \lb{F5} \ee

On the other hand, the variational problem can be solved by the use of Lagrange
multipliers for the
time-dependent constraints:
\be \delta\left(\Gamma[\Psi^R,\Psi^L]-\int_0^\infty
dt\,\,\left[\bh(t)\bdot\langle\Psi^L(t),\hBZ\Psi^R(t)\rangle
-\lambda(t)\langle\Psi^L(t),\Psi^R(t)\rangle\right]\right)=0,
\lb{F6} \ee
yielding
\be (\partial_t-\hL_\bh(t))|\Psi^R(t)\rangle= -\lambda(t)|\Psi^R(t)\rangle
\lb{F7} \ee
and
\be (\partial_t+\hL_\bh^\dagger(t))|\Psi^L(t)\rangle=
\lambda^*(t)|\Psi^L(t)\rangle. \lb{F8} \ee
In that case we see that
\be |\Omega^R(t)\rangle=\exp\left[\int_0^t
ds\,\,\lambda(s)\right]\cdot|\Psi^R(t)\rangle \lb{F9} \ee
and
\be |\Omega^L(t)\rangle=\exp\left[\int_t^\infty
ds\,\,\lambda^*(s)\right]\cdot|\Psi^L(t)\rangle. \lb{F10} \ee
Substituting these into the Eq.(\ref{F1}) and using the overlap constraint, we
obtain the expression for the
cumulant-generating function that
\begin{eqnarray}
W[\bh] & = & \int_0^\infty dt\,\, \lambda(t) \cr
    \, & = & \int dt\,\,\langle\Psi^L(t),
(-\partial_t+\hL+\bh(t)\bdot\hBZ)\Psi^R(t)\rangle. \lb{F11}
\end{eqnarray}
The last equation was obtained by applying $\Psi^L(t)$ on the left to
Eq.(\ref{F7}). Note that, indeed,
$\delta W[\bh]/\delta\bh(t)=\bZ(t)$ by a simple calculation:
\begin{eqnarray}
{{\delta W[\bh]}\over{\delta\bh(t)}} & = & \bZ(t)
    +\int_0^\infty ds\,\,\left[
\lambda(s)\langle{{\delta\Psi^L(s)}\over{\delta\bh(t)}},\Psi^R(s)\rangle
+\lambda(s)\langle\Psi^L(s),{{\delta\Psi^R(s)}\over{\delta\bh(t)}}
                \rangle\right] \cr
  \,& = & \bZ(t)+\int_0^\infty
ds\,\,\lambda(s){{\delta}\over{\delta\bh(t)}}\langle\Psi^L(s),\Psi^R(s)\rangle
\cr
  \,& = & \bZ(t). \lb{F12}
\end{eqnarray}
To obtain the first line we used Eqs.(\ref{F7}),(\ref{F8}) and to obtain the
last line we used again the overlap
condition. We therefore get directly from Eq.(\ref{F11}) that
\begin{eqnarray}
\Gamma[\bZ] & \equiv & \int_0^\infty dt\,\,\bh(t)\bdot\bZ(t)-W[\bh] \cr
          \,& = & \int_0^\infty
dt\,\,\langle\Psi^L(t),(\partial_t-\hL)\Psi^R(t)\rangle, \lb{F13}
\end{eqnarray}
as was claimed. $\,\,\,\,\Box$

\vspace{.1in}

\noindent As remarked above, the quantity
\be {\cal L}(t)\equiv  \langle\Psi^L(t),(\partial_t-\hL)\Psi^R(t)\rangle
\lb{Lagrfn} \ee
can be taken as a Lagrangian functional in  terms of which
$\Gamma=\int_{-\infty}^{+\infty} dt\,{\cal L}(t),$
i.e. a time-density for the effective action.

On the basis of this theorem a practical Rayleigh-Ritz scheme may be devised.
If the variation described in
the theorem is carried out within a finite-parameter ansatz such as
Eqs.(\ref{tdans}) for $\Psi^H,\,\,H=L,R$, then
the problem reduces to determining stationary points of a parametric action
\be \Gamma[\obalpha;\bh]\equiv \int_0^\infty
dt\,\left[\pi_i(\obalpha(t))\dot{\oalpha}_i(t)-\cH(\obalpha(t))
-\bh(t)\bdot\left(\bcZ(\obalpha(t))-\bZ(t)\right)+\lambda(t)
            \left(\cN(\obalpha(t))-1\right)\right], \lb{hparact} \ee
which incorporates the constraints by Lagrange multipliers $\bh(t),\lambda(t)$.
We have defined
\be \cZ_\mu(\obalpha)=\langle\oPsi^L(\obalpha),\HZ_\mu\oPsi^R(\obalpha)\rangle,
\lb{curlyZ} \ee
and
\be \cN(\obalpha)=\langle\oPsi^L(\obalpha),\oPsi^R(\obalpha)\rangle.
\lb{curlyN} \ee
As in the static case, the ansatz Eqs.(\ref{tdans}) may need to be
``augmented'' to allow for the fact that $\Psi^L(t)
\neq 1$ when $\bh(t)\neq 0$. We will consider here briefly just the simplest
situation, where $\oPsi^H=\oPsi^H(\obalpha^H),
\,\,H=L,R$, with $\oPsi^L$ given by Eq.(\ref{nw0}) and the $\obalpha^R$
parameters taken just to be the corresponding
moments $\obm$, as in Eqs.(\ref{nw10})-(\ref{nw13}). In this case, the
parametric action takes the form
\be \Gamma[\obm,\obalpha^L;\bh]\equiv \int_0^\infty
dt\,\left[\obalpha^L(t)\bdot\dot{\obm}(t)
-\obalpha^L(t)\bdot\obV(\obm(t),\bh(t))+\lambda(t)(\obalpha^L(t)
                            \bdot\obm(t)-1)\right], \lb{hparact'} \ee
neglecting some terms independent of the parameters being varied. The
corresponding Euler-Lagrange equations are
\be \dot{\obm}(t)=\obV(\obm(t),\bh(t))-\lambda(t)\obm(t), \lb{2ndEuLag} \ee
\be
\dot{\obalpha^L}(t)+\bA(\obm(t),\bh(t))\obalpha^L(t)=\lambda(t)\obalpha^L(t),
\lb{1stEuLag} \ee
\be \obalpha^L(t)\bdot\obm(t)=1, \lb{3rdEuLag} \ee
with the boundary conditions at initial and final times:
\be \obm(0)=\obm_0,\,\,\,\,\,\,\obalpha^L(+\infty)
                  =(1,\bz)\,\,\,\,\,\,\lambda(+\infty)=0. \lb{bc} \ee
These equations should be compared with their static counterparts,
Eqs.(\ref{nw10}),(\ref{nw13}).
For a specified $\bh(t)$, this {\em two-point boundary value problem} may be
solved numerically by standard methods:
see \cite{numrec}, Ch. 17. For small $\bh(t)$, the best numerical scheme is
probably the relaxation method,
because an exact solution is known for the system at $\bh(t)\equiv \bz$,
corresponding to a solution $\obm(t)$ of
the moment-closure dynamics with specified initial data $\obm(0)=\obm_0$ and to
$\obalpha^L(t)\equiv(1,\bz),
\lambda(t)\equiv 0$. This known solution for $\bh_0(t)\equiv \bz$ may then be
input as an initial guess into a relaxation
algorithm to find the solution with some small $\bh_1(t)$, and, iteratively, a
sequence of solutions with $\bh_k(t)$
of increasing magnitude constructed. In this way, the fluctuations around the
predicted dynamical trajectory $\obm(t)$
of the moment closure may be explored in the Rayleigh-Ritz method by varying
$\bh(t)$. The method then yields
an approximate effective action
\be \Gamma_*[\bZ]=\int_0^\infty dt\,
\left[\obalpha^L_*(t)\bdot\dot{\obm}_*(t)-\obalpha^L_*(t)
                      \bdot\obV(\obm_*(t))\right], \lb{appeffact} \ee
in which ${\obm}_*(t),\obalpha^L_*(t),\lambda_*(t)$ are solutions of the
initial-final value problem Eqs.(\ref{1stEuLag})
-(\ref{3rdEuLag}), with $\bh(t)$ selected so that
\be z_{*\mu}(t)\equiv \obalpha^L_*(t)\bdot \oBZ_{*\mu}(t) \lb{bigZ} \ee
equals the specified $z_\mu(t)$. We have defined
$\oBZ_{*\mu}(\obm)=\langle\HZ_\mu\overline{\bpsi}^L\rangle_\obm$.

Equivalently, the approximate action may be written as
\be \Gamma_*[\bZ]=\int_0^\infty
dt\,\left[\bZ_*(t)\bdot\bh(t)-\lambda_*(t)\right]. \lb{appeffact2} \ee
This can be compared with the approximate effective potential in Proposition 1.
If we define the approximate generating
functional $W_*[\bh]=\int_0^\infty dt\,\,\lambda_*(t)$, then it also follows as
in Proposition 1 that
\be {{\delta W_*[\bh]}\over{\delta h_\mu(t)}}=z_{*\mu}(t). \lb{timeFH} \ee
Thus, the approximate effective action from the Rayleigh-Ritz method,
Eq.(\ref{appeffact}) or Eq.(\ref{appeffact2}),
retains the Legendre transform structure of the true effective action. It is
not hard to derive from this fact that
\be \Gamma_*[\obZ_*]=0 \,\,\,\,\&\,\,\,\, {{\delta
\Gamma_*}\over{\delta\bZ(t)}}[\obZ_*]=\bz. \lb{Tconseq} \ee
where $\obZ_*(t)=\langle\bZ\rangle_{\obm(t)}$ is the expected value of $\bZ$ in
the PDF ansatz calculated
along the trajectory $\obm(t)$ of the moment-closure. Hence, the predicted
mean-history $\obZ_*(t)$ is guaranteed
to be a stationary point of $\Gamma_*[\bZ]$, but not necessarily a minimum
point.

\newpage

Recently, an alternative nonperturbative approximation to the nonequilibrium
effective action has been developed
by Crisanti \& Marconi \cite{CM}, via a dynamical Hartree approximation. While
the two approximation schemes
are similar in spirit, there are essential differences between them. We present
here no detailed comparison
of the two techniques. However, we believe it is a virtue of the present method
that it allows an approximation
of the effective action and effective potential within {\em any} PDF ansatz
that may be proposed. Furthermore,
it makes direct connection with the moment-closure equations which have been
traditionally used in nonequilbrium
statistical dynamics. We believe that the combination of flexibility to
incorporate intuitive guesses and transparency
of the physical interpretation should give the present method far-reaching
applications.

\vspace{.5in}

\noindent {\bf Acknowledgements}. I would like to renew my thanks to all
parties acknowledged in Ref.\cite{Eyi}.
I wish to give special thanks to the following: F. J. Alexander, whose
collaborative work on numerical implementation
of these ideas has helped to sharpen them considerably; B. Bayly, who
generously made available his own unpublished
work, which overlaps ours in many points; C. D. Levermore, who shared insights
from his related moment-closure methods
in kinetic theory and, as well, pointed out the Hamiltonian form of our
parametric equations, Eq.(\ref{parHameq});
and, finally, Y. Oono, who made many critical and useful suggestions during the
formative period of this variational
method.

\newpage

\section{Appendix: General Variational Equations}

\noindent The most general trial ansatz has the form
$\Psi^H=\Psi^H(\balpha,\balpha^H),\,H=L,R$ with $N^L=N+N^R$.
In this case, the parametric Hamiltonian is calculated as
\be \cH(\balpha,\balpha^R,\balpha^L)=\langle\Psi^L(\balpha,\balpha^L),
          \hL\Psi^R(\balpha,\balpha^R)\rangle. \lb{GHam} \ee
Correspondingly, the fixed point conditions are
\be {{\partial\cH}\over{\partial\alpha_i}}(\balpha,\balpha^R,\balpha^L),
    =\langle\psi^L_i(\balpha,\balpha^L),\hL\Psi^R(\balpha,\balpha^R)\rangle=0
\lb{FPT} \ee
and
\be {{\partial\cH}\over{\partial\alpha^R_i}}(\balpha,\balpha^R,\balpha^L)
    =\langle\Psi^L(\balpha,\balpha^L),\hL\psi^R_i(\balpha,\balpha^R)\rangle=0
\lb{FPR} \ee
and
\be {{\partial\cH}\over{\partial\alpha^L_i}}(\balpha,\balpha^R,\balpha^L)
=\langle{{\partial\Psi^L}\over{\partial\alpha_i}}
         (\balpha,\balpha^L),\hL\Psi^R(\balpha,\balpha^R)\rangle
+\langle\Psi^L(\balpha,\balpha^L),\hL{{\partial\Psi^R}
     \over{\partial\alpha_i}}(\balpha,\balpha^R)\rangle=0 \lb{FPL} \ee
with $\psi^H_i={{\partial\Psi^H}\over{\partial\alpha_i^H}},\,\,H=R,L$. Within
the same ansatz, the parametric evolution
equations have the form
\be \{\alpha_i,\alpha_j\}\dot{\alpha}_j+\{\alpha_i,\alpha_j^R\}
  \dot{\alpha}^R_j+\{\alpha_i,\alpha_j^L\}\dot{\alpha}^L_j
={{\partial\cH}\over{\partial\alpha_i}}(\balpha,\balpha^R,\balpha^L),
\lb{parmeq} \ee
and
\be
\{\alpha_i^R,\alpha_j\}\dot{\alpha}_j+\{\alpha_i^R,\alpha_j^L\}\dot{\alpha}^L_j
={{\partial\cH}\over{\partial\alpha^R_i}}(\balpha,\balpha^R,\balpha^L),
\lb{parmeqR} \ee
and
\be
\{\alpha_i^L,\alpha_j\}\dot{\alpha}_j+\{\alpha_i^L,\alpha^R_j\}\dot{\alpha}^R_j
={{\partial\cH}\over{\partial\alpha^L_i}}(\balpha,\balpha^R,\balpha^L).
\lb{parmeqL} \ee
The most general ansatz of any obvious utility is that given in
Eq.(\ref{ansl*}):
\be \Psi^L= 1+\sum_{i=1}^N\alpha_i^L\psi_i^L(\balpha)
\,\,\,\,\&\,\,\,\,\Psi^R=\Psi^R(\balpha). \lb{ansl**} \ee
This may be thought to correspond to the previous ansatz with $N^R=0,N^L=N$ and
with a linear dependence of $\Psi^L$
on the $\balpha^L$. For this case, the parametric Hamiltonian is
\be \cH(\balpha,\balpha^L)=\sum_{i=1}^N\alpha_i^L V_i(\balpha) \lb{parham*} \ee
with $V_i(\balpha)=\langle\hL^\dagger\psi_i^L(\balpha)\rangle_\balpha$ the
dynamical vector field in the parameter
space, as in Eq.(\ref{gendynvec}). The fixed point conditions are simply
\be V_i(\balpha)= 0 \lb{FPL*} \ee
and
\be \alpha_j^L {{\partial V_j}\over{\partial\alpha_i}}(\balpha)=0 \lb{FP*} \ee
for $i=1,...,N$. When the stability matrix at a fixed point $\balpha_*$ of the
first equation (\ref{FPL*}) is
non-singular, ${\rm det}\left[{{\partial
\bV}\over{\partial\balpha}}(\balpha_*)\right]\neq 0$, then the only
solution of the second equation is $\balpha^L=\bz$. The parametric evolution
equations within the same ansatz are
\be \{\alpha_i,\alpha_j\}\dot{\alpha}_j+\{\alpha_i^L,\alpha_j\}\dot{\alpha}_j=
V_i(\balpha) \lb{parmeqL*} \ee
and
\be \{\alpha_i,\alpha_j^L\}\dot{\alpha}_j^L=\alpha_j^L {{\partial
V_j}\over{\partial\alpha_i}}(\balpha), \lb{parmeq*} \ee
where the Lagrange brackets are
\be \{\alpha_i,\alpha_j\}
=\sum_{k=1}^N\alpha_k^L\left[\langle{{\partial\psi_k^L}
     \over{\alpha_i}}(\balpha),\psi_j^R(\balpha)\rangle
-\langle{{\partial\psi_k^L}\over{\alpha_j}}(\balpha),
\psi_i^R(\balpha)\rangle\right] \lb{Lagbra*} \ee
and
\be \{\alpha_i^L,\alpha_j\}=\langle\psi_i^L(\balpha),\psi_j^R(\balpha)\rangle
\lb{LagbraL*} \ee
with now $\psi^R_i\equiv {{\partial\Psi^R}\over{\partial\alpha_i}}$. The second
equation clearly has the
constant solution $\balpha^L(t)\equiv \bz$. The first equation then has the
same form as Eq.(\ref{sparmeqL*})
in the text. It is also identical with Eq.(2.11) in the work of Bayly
\cite{Bay}, but here derived by the variational
method.

\end{document}